\documentclass[twocolumn]{aastex63}

\usepackage{color}
 \received{Feb 16, 2022}
 \revised{Mar 22, 2022}
 \accepted{Mar 30, 2022}
 \submitjournal{ApJS}

\shorttitle{Validations and corrections of the SFD and \emph{Planck} maps}
\shortauthors{Sun, Yuan \& Chen}
\graphicspath{{./}{figures/}}

\begin{document}

\title{Validations and corrections of the SFD and \emph{Planck} reddening maps based on LAMOST and \emph{Gaia} data
}

\author[0000-0001-6561-9443]{Yang Sun}
\affiliation{Department of Astronomy,
  Beijing Normal University,
  Beijing 100875, China}
\affiliation{Steward Observatory,
University of Arizona, 
Tucson, AZ 85721, USA}

\author[0000-0003-2471-2363]{Haibo Yuan}
\affiliation{Department of Astronomy,
  Beijing Normal University,
  Beijing 100875, China}
\email{yuanhb@bnu.edu.cn}

\author[0000-0003-2472-4903]{Bingqiu Chen}
\affiliation{South-Western Institute for Astronomy Research, Yunnan University, Kunming 650500, China}

\begin{abstract}

  Precise correction of dust reddening is fundamental to obtain the intrinsic parameters of celestial objects. The Schlegel et al. (SFD) and the \emph{Planck} 2D extinction maps are widely used for the reddening correction. In this work, using accurate reddening determinations of about two million stars from the Large Sky Area Multi-Object Fiber Spectroscopic Telescope (LAMOST) data release 5 (DR5) and \emph{Gaia} DR2, we check and calibrate the SFD and \emph{Planck} maps in the middle and high Galactic latitudes. The maps show similar precision in reddening correction. We find small yet significant spatially dependent biases for the four maps, which are similar between the SFD and Planck2014-R maps, and between the Planck2014-Tau and Planck2019-Tau maps. The biases show a clear dependence on the dust temperature and extinction for the SFD and Planck2014-R maps. While those of the Planck2014-Tau and Planck2019-Tau maps have a weak dependence on the dust temperature, they both strongly depend on the dust spectral index. Finally, we present corrections of the SFD and \emph{Planck} extinction maps within the LAMOST footprint, along with empirical relations for corrections outside the LAMOST footprint. 
  Our results provide important clues for the further improvement of the Galactic all-sky extinction maps and 
  lay an significant foundation for the accurate extinction correction in the era of precision astronomy.

\end{abstract}



\section{Introduction} \label{sec:intro}

Interstellar dust is formed from the condensation of heavy elements which are expelled into the space by stellar winds or explosions. In the ultraviolet (UV), optical, and near-infrared (NIR) bands, dust absorbs and scatters the light passing through it. Therefore, correction for the effect of dust extinction is fundamental to reveal the intrinsic properties of observed objects. The existing extinction maps serve as a straight and convenient tool to correct for extinction. 

The two-dimensional (2D) extinction map of \citet[hereafter SFD]{schlegel_application_1998} is most widely used for dust corrections. The SFD map is based on the Infrared Astronomical Satellite (IRAS) 100\,$\mu$m and the Dust Infrared Background Experiment (DIRBE) 100 and 240\,$\mu$m data. The far-infrared (FIR) emission is firstly modeled by a modified blackbody to obtain the dust temperature, which is then used to determine the dust column density and finally transferred to reddening by calibrating against 389 elliptical galaxies. Unfortunately, the map suffers from moderate systematic uncertainties as the dust temperatures have a low spatial resolution as well as the fact that the cosmic infrared background (CIB) and the zodiacal light have not been fully removed.

Many works have been carried out to validate and calibrate the SFD map. 
By comparing the surface number density and the average colors of galaxies from the Sloan Digital Sky Survey (SDSS; \citealt{2000AJ....120.1579Y}) in the regions with different extinctions, \citet{yahata_effect_2007} indicated that, toward low-reddening lines of sight, the results show the opposite of the effect of Galactic dust that the number density of galaxies increases with $E(B-V)$ regardless of extinction correction with the SFD map. These unexpected effects were explained by the pollution of the sky brightness by extragalactic infrared emission of order 0.01 magnitudes in the SFD emission map.
Using 151,637 passively evolving galaxies from the SDSS, \citet{peek_correction_2010} have presented corrections to the SFD map in the high Galactic latitude region. They found that the SFD map under-predicts extinction in regions of low dust temperature. The result is further confirmed by a
large-scale reddening map from neutral hydrogen emission (\citealt{lenz_new_2017}).
By measuring the reddening of a sample of over 50, 000 QSOs from the SDSS DR7, \citet{wolf_milky_2014} concluded that there is a non-linearity in the SFD map (see also Figure 10 of \citealt{li_galactic_2017}), which may be caused by pollution from unresolved extragalactic infrared emission.
In addition, \citet{schlafly_blue_2010}, \citet{schlafly_measuring_2011} and \citet{yuan_empirical_2013} measured the stellar reddening and found that the SFD extinction map overestimates $E(B-V)$ by about 14 percent. 


Although many works have checked the errors of the SFD map, 
further investigations are still needed due to the limitations of previous works.
The limitations include large statistical uncertainties caused by small sample sizes, 
systematic calibration errors in the photometric data used, and invalid assumptions adopted.
Thanks to the Large Sky Area Multi-Object Fiber Spectroscopic Telescope (LAMOST; \citealt{cui_large_2012}) and \emph{Gaia} (\citealt{gaia_collaboration_gaia_2016}),  reddening values of an enormous number of stars can be estimated with high accuracy. 
The LAMOST surveys (\citealt{zhao_lamost_2012}; \citealt{deng_lamost_2012}; \citealt{liu2014proc}) have obtained high-quality spectra and precise atmospheric parameters including effective temperature, surface gravity, and metallicity for millions of stars. Their reddening values $E(G_{\rm BP}-G_{\rm RP})$ can be accurately measured from the star-pair method \citep{yuan_empirical_2013} by combining with the
\emph{Gaia} photometric data. 

In this work, we first fit empirical temperature- and reddening-dependent coefficient (R)
to convert $E(G_{\rm BP}-G_{\rm RP})$ into $E(B-V)$. Then by comparing with the SFD dust map, we produce a correction map for it and investigate the potential error sources. Since \citet{planck_collaboration_planck_2014-6} (hereafter Planck2014) applied a similar method as SFD to build a new extinction map (hereafter Planck2014-R) and provided another one based on 353 GHz absorption (hereafter Planck2014-Tau), we inspect and correct them at the same time. The map from \citet{irfan_determining_2019} (hereafter Planck2019-Tau), which is an update of Planck2014-Tau, is also considered. 

The paper is organized as follows. We describe our data and the method of calculating $E(B-V)$ in Section \ref{sec:data}. The details of the four extinction maps and the way to build the correction factor \emph{k} maps are introduced in Section \ref{sec:correction}. We present the final correction factor \emph{k} maps and discuss potential error sources of the extinction maps in Section \ref{sec:ReandDIS}. We summarize in Section \ref{sec:Sum}.

\section{Accurate reddening estimation from LAMOST and \emph{Gaia} data}
\label{sec:data}

We use a star-pair method (\citealt{yuan_empirical_2013}) to measure reddening values of individual stars.
The method assumes that stars of the same stellar atmospheric parameters should have the same intrinsic colors.
The method requires a control sample of stars that are unreddened or of very low reddening. We can then obtain the intrinsic colors of a reddened star from its pairs out of the control sample having the same atmospheric parameters. In this work, we calculate values of $E(G_{\rm BP}-G_{\rm RP})$ using the same star-pair algorithm of \citet{yuan_lamost_2015}.

\subsection{LAMOST and Gaia data}
\label{subsec:LAMOSTandGaia}

The stellar parameters used in this analysis are from the LAMOST Data Release 5 (DR5; \citealt{luo_first_2015}). The LAMOST spectra have a resolution of about $R$ $~$ 1800 and cover a wavelength range of 3700 - 9000\AA. The LAMOST DR5 has provided stellar atmospheric parameters (effective temperatures $T_{\mathrm{eff}}$, surface gravities $\log \mathrm{g}$, and metallicities $\mathrm{[Fe/H]}$) 
for over 5 million stars, with the LAMOST Stellar Parameter Pipeline (LASP; \citealt{wu_automatic_2011}). 
The typical uncertainties of the stellar parameters are about 110\,K, 0.2\,dex and 0.15\,dex for $T_{\mathrm{eff}}$, $\log \mathrm{g}$ and $\mathrm{[Fe/H]}$, 
respectively (\citealt{luo_first_2015}). Note the internal uncertainties are significantly lower (e.g., \citealt{niu_correction_2021}). 

To obtain accurate reddening values of LAMOST stars, we adopt the photometric data from the \emph{Gaia} DR2 (\citealt{gaia_collaboration_gaia_2018}). The \emph{Gaia} DR2 provides high precision $G$ magnitudes of $\sim$ 1.7 billion sources and $G_{\rm BP}$ and $G_{\rm RP}$ photometric measurements of $\sim$ 1.4 billion targets. The \emph{Gaia} $G$, $G_{\rm BP}$, and $G_{\rm RP}$ filters cover wavelength ranges of 330 - 1050, 330 - 680, and 630 - 1050\,nm, respectively. The calibration
uncertainties are 2, 5, and 3\,mmag for $G$, $G_{\rm BP}$, and $G_{\rm RP}$, respectively (\citealt{evans_gaia_2018}).

To validate and correct 2D reddening maps, we only use stars at medium and high Galactic latitudes, where dust layers can be treated as thin screens and the reddening values of the observed stars are the same as those at infinite distances. Stars from the LAMOST DR5 are first cross-matched with the \emph{Gaia} DR2. We then select stars with the following criteria: $\rm{parallax} > 0$, Galactic latitude $|b|>20 $\arcdeg, distance to the Galactic plane $|Z| > 200\,\mathrm{pc}$, effective temperature $T_{\mathrm{eff}} \in [4000,8000]\,\mathrm{K}$, LAMOST spectral signal-to-noise ratio (SNR) at $g$-band ${\rm {SNR}_{g}} > 20$. This yields 2,083,741 sources as our sample. 

\subsection{Reddening determination}
\label{subsec:calEBV}

As we mentioned above, the star-pair method is used to estimate the reddening values for our sample stars.
For a given sample star, its control stars are selected via the following criteria: $\Delta{T_\mathrm{eff}}<T_\mathrm{eff} * {(0.00003 * T_\mathrm{eff})}^{2}\,\mathrm{K}$, $\Delta{\log \mathrm{g}}<0.5\, \mathrm{dex}$, $\Delta{\mathrm{[Fe/H]}}<0.3\, \mathrm{dex}$, distance to the Galactic plane $|Z| > 1\,\mathrm{kpc}$ and extinction value given by the SFD map $E(B-V)_{SFD} < 0.02\,\mathrm{mag}$. Then by using the same star-pair algorithm described in \citet{yuan_lamost_2015}, we obtain $E(G_{\rm BP}-G_{\rm RP})$ and $E(G_{\rm BP}-G)$ for our sample. 
Stars of resulted reddening $E(G_{\rm BP}-G_{\rm RP})<-0.05$ mag are excluded. This leads to 2,067,805 stars in our sample, which we denote as `LAMOST sample' in this work.

Reddening coefficient R is usually used to convert reddening values of a given color to $E(B-V)$. 
The stellar extinction depends on the convolution of stellar spectral energy distribution (SED) and the extinction curve. Therefore, the reddening coefficient R is physically related to the $T_{\mathrm{eff}}$ and the $E(B-V)$.
This effect is particularly strong for the very broad \emph{Gaia} passbands. Therefore, we 
use binary quadratic functions to fit $R_{G_{\rm BP}-G_{\rm RP}}$ and $R_{G_{\rm BP}-G}$ as functions of 
$T_{\mathrm{eff}}$ and $E(B-V)$. 
A sample of stars is collected via the following criteria:
\begin{itemize}
\item{$|b| > 15$\arcdeg;}
\item{$|Z|>300\,\mathrm{pc}$;}
\item{If $T_{\mathrm{eff}} > 4500\,\mathrm{K}$, $\mathrm{{SNR}_{g}} >15$.
Otherwise, $\mathrm{{SNR}_{g}}>20$;}
\item{$E(B-V)>0.05\,\mathrm{mag}$.}
\end{itemize}
The selection criteria are different from those for selecting stars to correct reddening maps mentioned in Section \ref{subsec:LAMOSTandGaia}. To make the fitting more reliable, here we exclude stars in very low reddening regions, and restrict $\mathrm{{SNR}_{g}}$ more tightly. This yields about 700,000 sources for fitting. The exact number varies slightly for different extinction maps.

In order to build grids, $E(B-V)$ values are equally divided into 8 bins from 0 to 0.8 mag, 
and $T_{\mathrm{eff}}$ are also equally divided into 8 bins from 4000 to 8000\,K. Then for each grid, 
the medians of $E(B-V)$, $T_{\mathrm{eff}}$, and R are estimated. The error of R is also calculated by the following formula:

\begin{equation}     
R_{err} = \frac{std(R)}{\sqrt{N}},
\end{equation}  
where N is the number of sources in the grid. The grids with N $<$ 10 or $R_{err}$ $>$ 0.03 are discarded in the fitting. 

\begin{figure*}
  \gridline{\fig{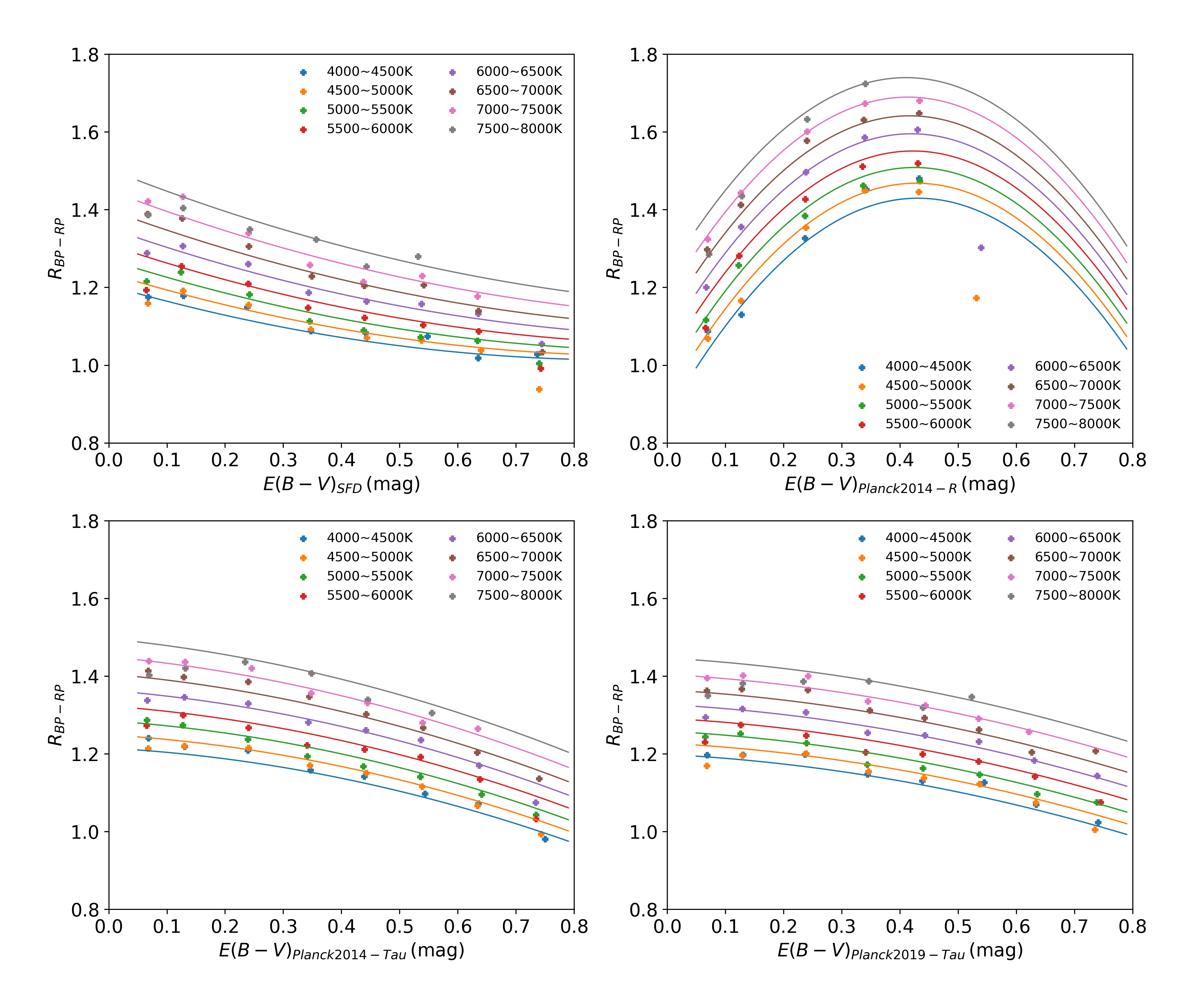}{0.9\textwidth}{}
            }
    \caption{Reddening coefficients of ${G_{\rm BP}-G_{\rm RP}}$ color as a function of reddening for different reddening maps. 
    In each panel, the dots and lines are measured and fitted values, respectively. Different colors denote different 
    temperature ranges of stars used. For the SFD, Planck2014-Tau, and Planck2019-Tau maps, only dots of $E(B-V) < 0.7$ mag 
    are used. For the Planck-R map, only dots of $E(B-V) < 0.5$ mag are used. Note that for the SFD, Planck2014-Tau, and Planck2019-Tau maps, the higher stellar temperatures, the lower reddening values, the larger the reddening coefficients.
    While for the Planck-R map, the reddening coefficients are larger for higher reddening values when $E(B-V) < 0.45$ mag.  
    }
    \label{Fig1:R_fitting_BPRP}
\end{figure*}

\begin{figure*}
  \gridline{\fig{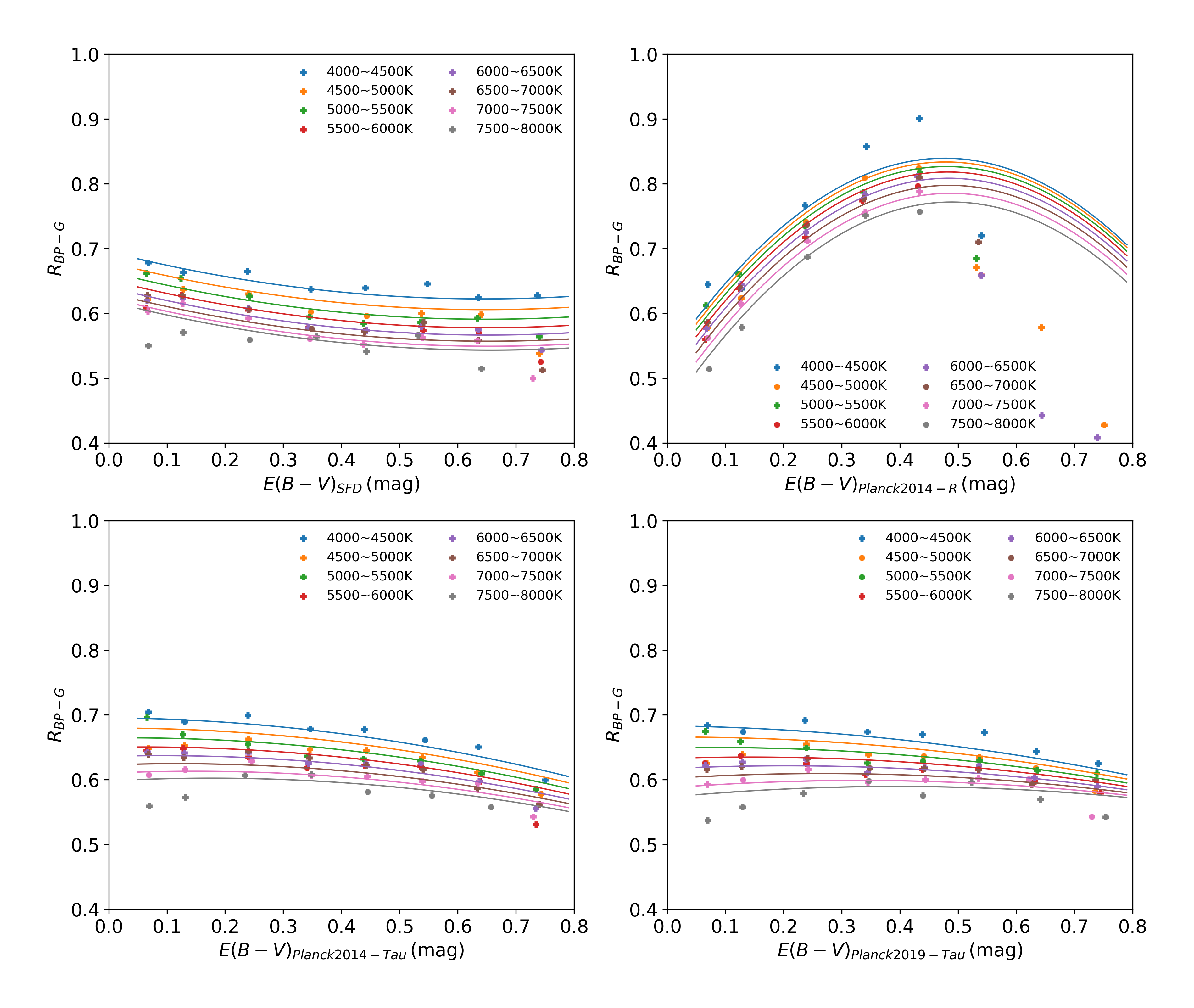}{0.9\textwidth}{}
            }
  \caption{Same to Figure\,\ref{Fig1:R_fitting_BPRP} but for the ${G_{\rm BP}-G}$ color. Note that for the SFD, Planck2014-Tau, and Planck2019-Tau maps, the lower
  stellar temperatures, the lower reddening values, the larger the reddening coefficients.}
  \label{Fig2:R_fitting_BPG}
\end{figure*}

We fit the relations for the four maps separately, considering their different systematics in $E(B-V)$. 
As to be mentioned in Section \ref{subsec:dustmap}, the  Planck2014-R map is unreliable in high extinction regions.
Therefore, for this map we only choose grids whose $E(B-V)$ values are in the range of $[0.05,0.5]$ mag,  
and in the range of $[0.05,0.7]$ mag for other maps. 
The fitting coefficients are listed in Table \ref{Tab1:R_BPRPandBPG}. The results are plotted 
in Figure \ref{Fig1:R_fitting_BPRP} and Figure \ref{Fig2:R_fitting_BPG}.
Note that the trend with $E(B-V)$ is different for the Planck2014-R map. For the other three maps, 
$R_{G_{\rm BP}-G_{\rm RP}}$ and $R_{G_{\rm BP}-G}$ decrease as $E(B-V)$ increases. 
For the Planck2014-R map, $R_{G_{\rm BP}-G_{\rm RP}}$ and $R_{G_{\rm BP}-G}$ increase as $E(B-V)$ increases 
at $E(B-V) < 0.4$\,mag. It is because that the Planck2014-R map tends to underestimate reddening in high 
extinction regions.

Via the above relations, the $E(B-V)_{\rm LAMOST}$ values can be computed:

\begin{equation}     
  E(B-V)_{\rm LAMOST} = \frac{E(G_{BP}-G_{RP})}{R_{G_{BP}-G_{RP}}}
\end{equation} 
The typical errors of $E(B-V)_{\rm LAMOST}$ for individual stars are around 0.01--0.02 mag.

\section{Corrections of reddening maps}
\label{sec:correction}
\subsection{The SFD and Planck maps} \label{subsec:dustmap}

The SFD map traces dust reddening via thermal emission in the far-infrared based on the IRAS data. The IRAS 100\,$\mu$m map was calibrated to match the COBE/DIRBE data. The zodiacal light and CIB contaminations 
were removed. The DIRBE 100 and 240\,$\mu$m data were then used to estimate the dust temperature at a spatial 
resolution of 1.3$^\circ$ and to transform the IRAS 100\,$\mu$m flux to dust optical depth. 
By using a sample of 389 elliptical galaxies, the dust map was finally normalized to $E(B-V)$ reddening map. 
The SFD reddening map has a spatial resolution of 6.1\,arcmin.

The Planck2014-R and Planck2014-Tau maps are based on the \emph{Planck} 350, 550, and 850 $\mu$m data from the HFI 2013 delivery maps (\citealt{planck_collaboration_planck_2014-6}) and the IRAS 100 $\mu$m data. Dust parameters, including the dust optical depth $\tau_{353}$, dust temperature $T_{\rm obs}$, and spectral index of the dust emission $\beta_{\rm obs}$, were obtained by the $\chi^2$ fit of the observed dust SED with a modified blackbody (MBB) model. The dust radiance $\mathcal{R}$ was also calculated by integrating the MBB fit. Finally, Planck2014-R and Planck2014-Tau reddening maps were respectively obtained by transforming $\mathcal{R}$ and $\tau_{353}$ to $E(B-V)$ using extinction measurements of SDSS quasars. Both maps have a resolution of 5\,arcmin.
Planck2014 recommended that their Planck2014-R reddening map should only be used to estimate reddening in lines of sight where $E(B-V)<$ 0.3 mag. However, to make a more robust comparison, we slightly enlarged this limit by adopting sightlines of reddening $E(B-V) < $0.5 mag. Finally, Planck2014 indicated an offset of $-$0.003 mag for the SFD map.
We thus corrected the SFD maps by adding the offset.

Based on the \emph{Planck} Release 2 353, 545, and 857 GHz maps as well as the IRAS 100\,$\mu$m data, \citet{irfan_determining_2019} determined the MBB model parameters of thermal dust emission by a new sparsity-based, parametric method and presented the Planck2019-Tau map. By adopting the new method, they are able to produce full-resolution MBB parameter maps without smoothing and account for the CIB without removing thermal dust emission through over-smoothing. We use the same coefficient for the Planck2014-Tau map to convert $\tau_{353}$ to $E(B-V)$. The Planck2019-Tau map has a resolution of 5\,arcmin.

\subsection{Correction Factor}
\label{subsec:k}

\begin{figure*}
  \gridline{\fig{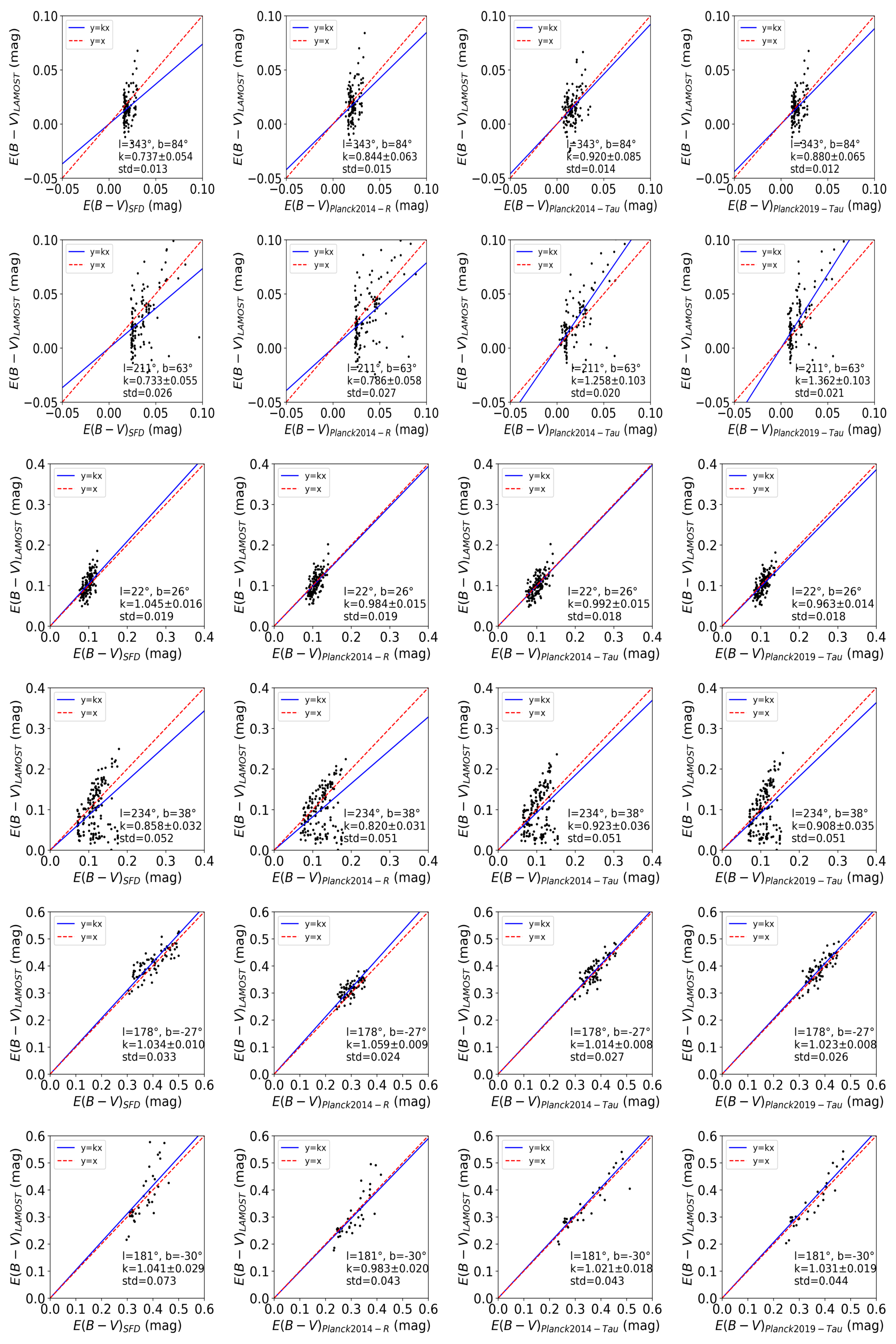}{0.8\textwidth}{}             }
\caption{LAMOST reddening against those from different reddening maps for six selected sight-lines of  
$E(B-V)$  about 0.02 (two top rows), 0.1 (two middle rows), and 0.4 mag (two bottom rows), respectively. 
From left to right are for the SFD, Planck2014-R, Planck2014-Tau, and Planck2019-Tau maps, respectively. 
For each panel, the blue line represents the final $k$, the red dashed line denotes
line of equality. The spatial position, estimated $k$ value, and standard deviation are also labeled.
The 1st, 3rd, and 5th rows have typical standard deviations, 
while the 2nd, 4th, and 6th rows show examples of abnormal standard deviations.}
\label{Fig3:k_sample_pixel}
\end{figure*}

To validate and correct the SFD and \emph{Planck} maps, we first obtain the $E(B-V)_{\rm map}$ values for the individual stars in the LAMOST sample from the SFD and \emph{Planck} maps, and then compare these reddening values to those derived from the star-pair method $E(B-V)_{\rm LAMOST}$. Sample stars are then divided into different subfields (pixels) by the HEALPix grid \citep{gorski_healpix_2005} at {\tt\string Nside} = 64 (with a resolution of about 1\degr). There are typically 100 stars in one pixel. For each pixel, we assume that 
$E(B-V)_{\rm LAMOST} = k \times E(B-V)_{\rm map}$,  
where $k$ is the correction factor and is estimated by averaging the ratios of $E(B-V)_{\rm LAMOST}$ and $E(B-V)_{\rm map}$ after 3$\sigma$ clipping. 
Its uncertainty is derived by,
\begin{equation}     
  k_{\rm err} = \frac{Std(\frac{E(B-V)_{\rm LAMOST}-E(B-V)_{\rm fit}}{E(B-V)_{\rm map}})}{\sqrt{N}},
\end{equation} 
where $E(B-V)_{\rm fit} = k \times E(B-V)_{\rm map}$ and N is the total number of sources for a given pixel. 
Figure \ref{Fig3:k_sample_pixel} shows the comparisons of $E(B-V)_{\rm LAMOST}$ and $E(B-V)_{\rm map}$ in
six selected sight-lines. For a given sightline, the standard deviations between different maps 
are similar. The typical standard deviations are between 0.01 -- 0.03\,mag, increasing 
towards high extinction sight-lines.

\section{Result and Discussion}
\label{sec:ReandDIS}

\subsection{Removing bad pixels}
\label{subsec:badpix}

\begin{figure*}
  \gridline{\fig{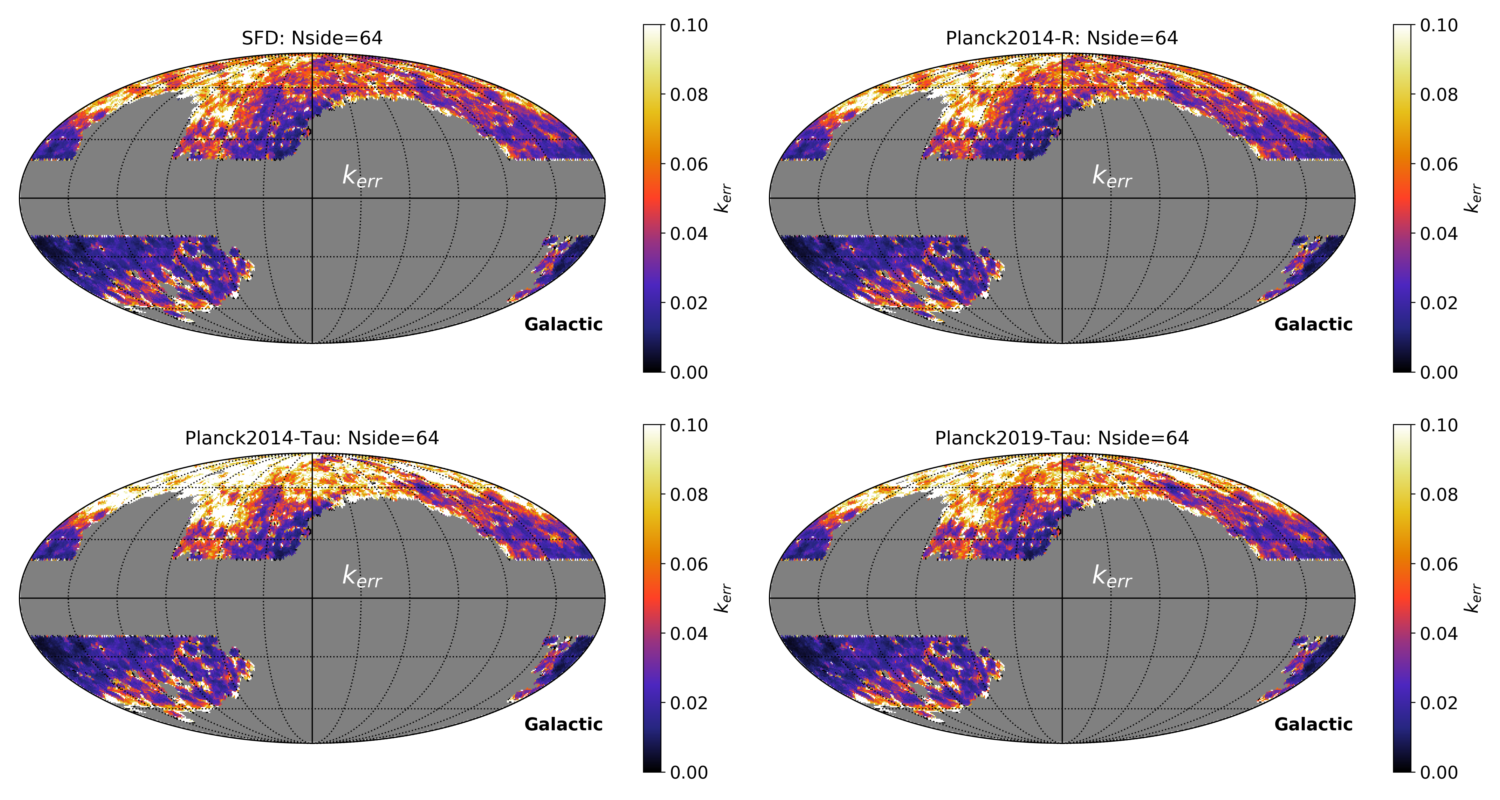}{0.7\textwidth}{}
            }
  \caption{Spatial distributions of correction factor errors $k_{err}$ in Galactic coordinates 
  for different reddening maps.}
  \label{Fig4:kerror_original}
\end{figure*}

Figure \ref{Fig4:kerror_original} shows the spatial variations of $k_{\rm err}$ for the SFD and \emph{Planck} reddening maps. 
Excluding pixels in the very low extinction regions of the maps, 
most $k_{\rm err}$ values are less than 0.1, only a few pixels have $k_{\rm err} > 0.1$ 
due to their small numbers of stars or large deviations between ${E(B-V)}_\mathrm{LAMOST}$ and ${E(B-V)}_{fit}$.

\begin{figure*}
  \gridline{\fig{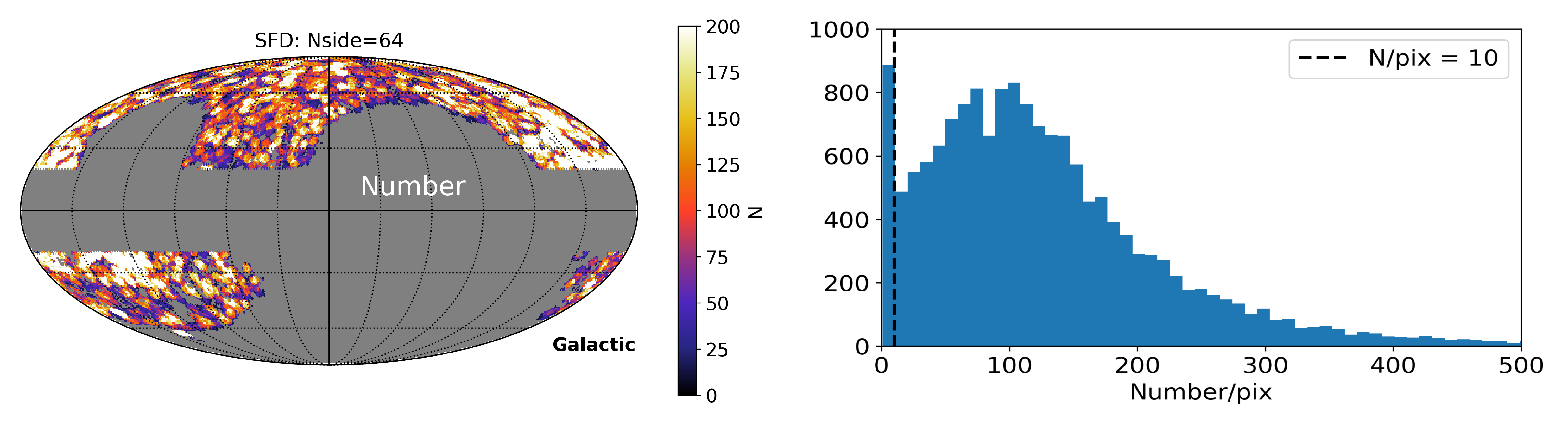}{0.9\textwidth}{}
            }
  \caption{Left: spatial distribution of the number of stars in each pixel for the SFD map. Note that due to independent 3$\sigma$ clipping for each extinction map, the numbers of stars in each pixel for the four extinction maps may be slightly different. Right: histogram distribution of the number of remaining sources in each pixel after 3$\sigma$ clipping. The vertical black line 
  indicates $N/pix=10$.}
  \label{Fig5:number}
\end{figure*}

Figure \ref{Fig5:number} plots the spatial and histogram distributions 
of star numbers in the individual pixels for the SFD map as an example. There are about 100 stars 
for a typical pixel. Only 887 pixels have less than 10 stars. The results for the three Planck maps are similar. 
We thus exclude these pixels in the following analyses. 

\begin{figure*}
  \gridline{\fig{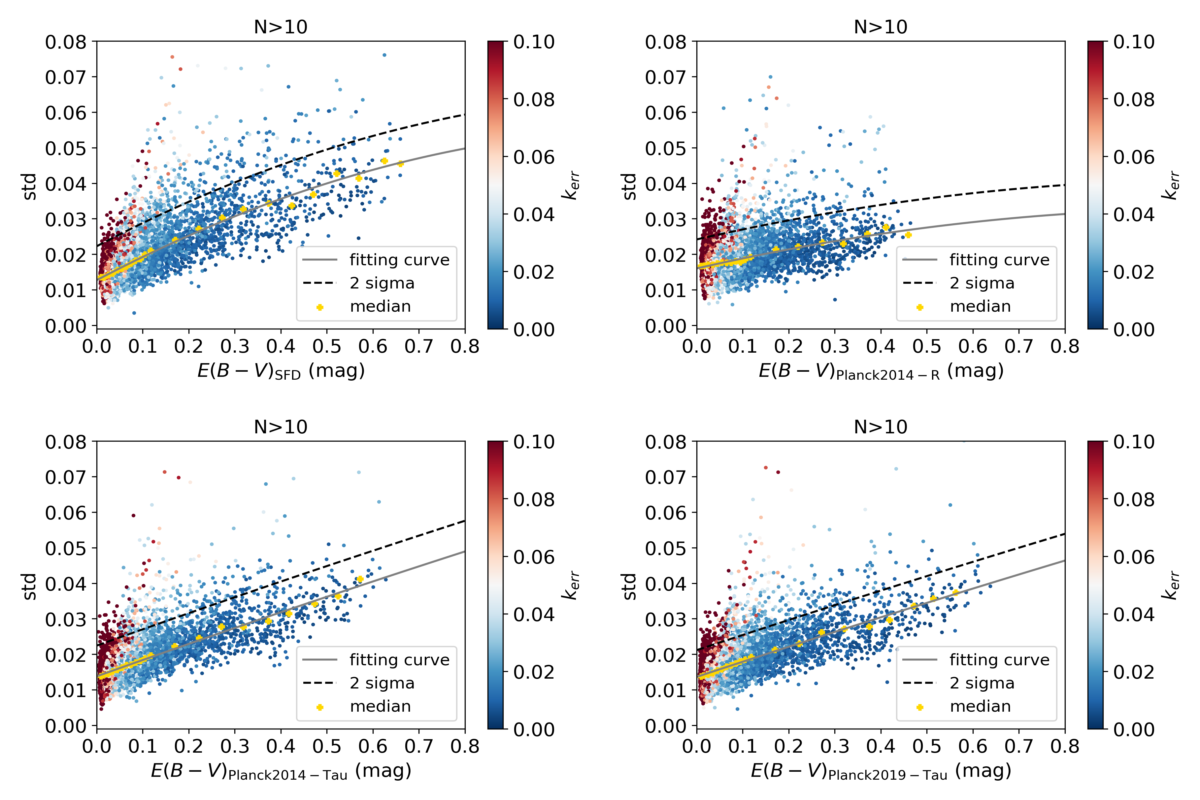}{0.8\textwidth}{}
            }
  \caption{$Std_{fit}$ of the SFD, Planck2014-R, Planck2014-Tau, and Planck2019-Tau maps as a function of reddening. The color of the dots indicates $k_{err}$. In each panel, the dots are divided into different bins. The bin width is 
  0.01/0.1 mag when the extinction value is smaller/larger than 0.1 mag. For each panel, the yellow crosses represent the median values, the grey line represents a second-order polynomial fitting to the yellow crosses, 
  the black dash line is obtained by shifting the grey line up by 2-$\sigma$ of the fitting. 
  The dots above the black dash lines are excluded.}
  \label{Fig6:stdcut}
\end{figure*}

Figure \ref{Fig6:stdcut} shows ${\rm Std}_{\rm fit}$ against $E(B-V)_{\rm map}$ for the 
SFD and \emph{Planck} reddening maps, where ${\rm Std}_{\rm fit}$ is the standard deviation of the reddening fit residuals ${\rm Std}_{\rm fit} = Std(E(B-V)_{\rm LAMOST}-E(B-V)_{\rm fit})$. 
Values of $k_{\rm err}$ are also color-coded.
Only pixels having over 10 stars are plotted. ${\rm Std}_{\rm fit}$ increases with $E(B-V)_{\rm map}$. 
Second-order polynomials are applied to fit the binned median values. The pixels that 
have ${\rm Std}_{\rm fit}$ values larger than the fitted curves by 2$\sigma$ deviation are
excluded. 
There are about 400 pixels in each panel of Figure \ref{Fig6:stdcut}  that have very large values of ${\rm Std}_{\rm fit}$ and consequently $k_{\rm err}$ $>$ 0.1. 
Note not all pixels eliminated from the four extinction maps are the same. A total of 739 pixels
are discarded from the four extinction maps. 

\begin{figure*}
  \gridline{\fig{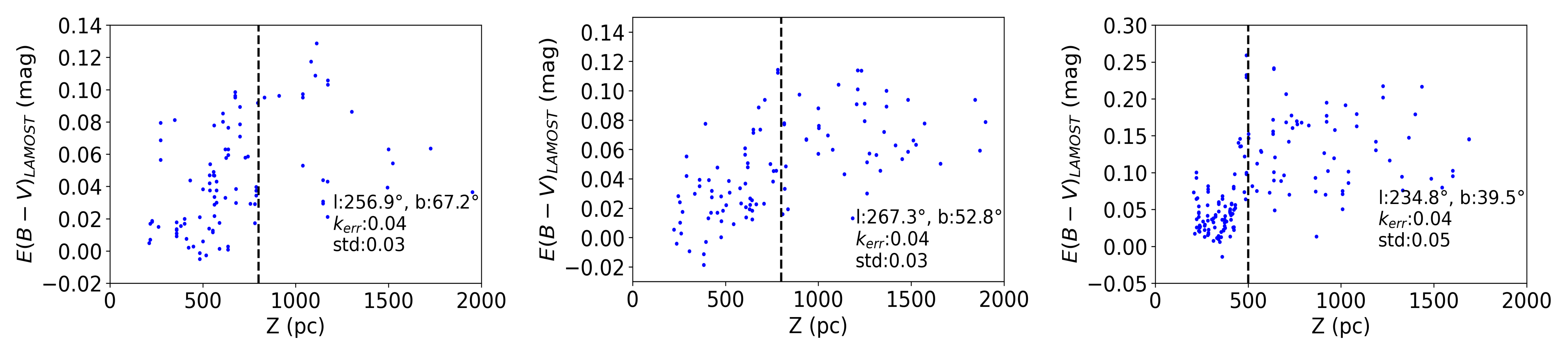}{1.0\textwidth}{}
            }
  \caption{Three examples of sight-lines that show large $Std_{fit}$ values (dots above the black dashed lines in Figure~\ref{Fig6:stdcut}). 
  Dust clouds far above the Galactic disk ($Z \sim$500 pc) are found as jumps of $k$ in the panels. }
  \label{Fig7:samples_fgdust}
\end{figure*}

\begin{figure*}
  \gridline{\fig{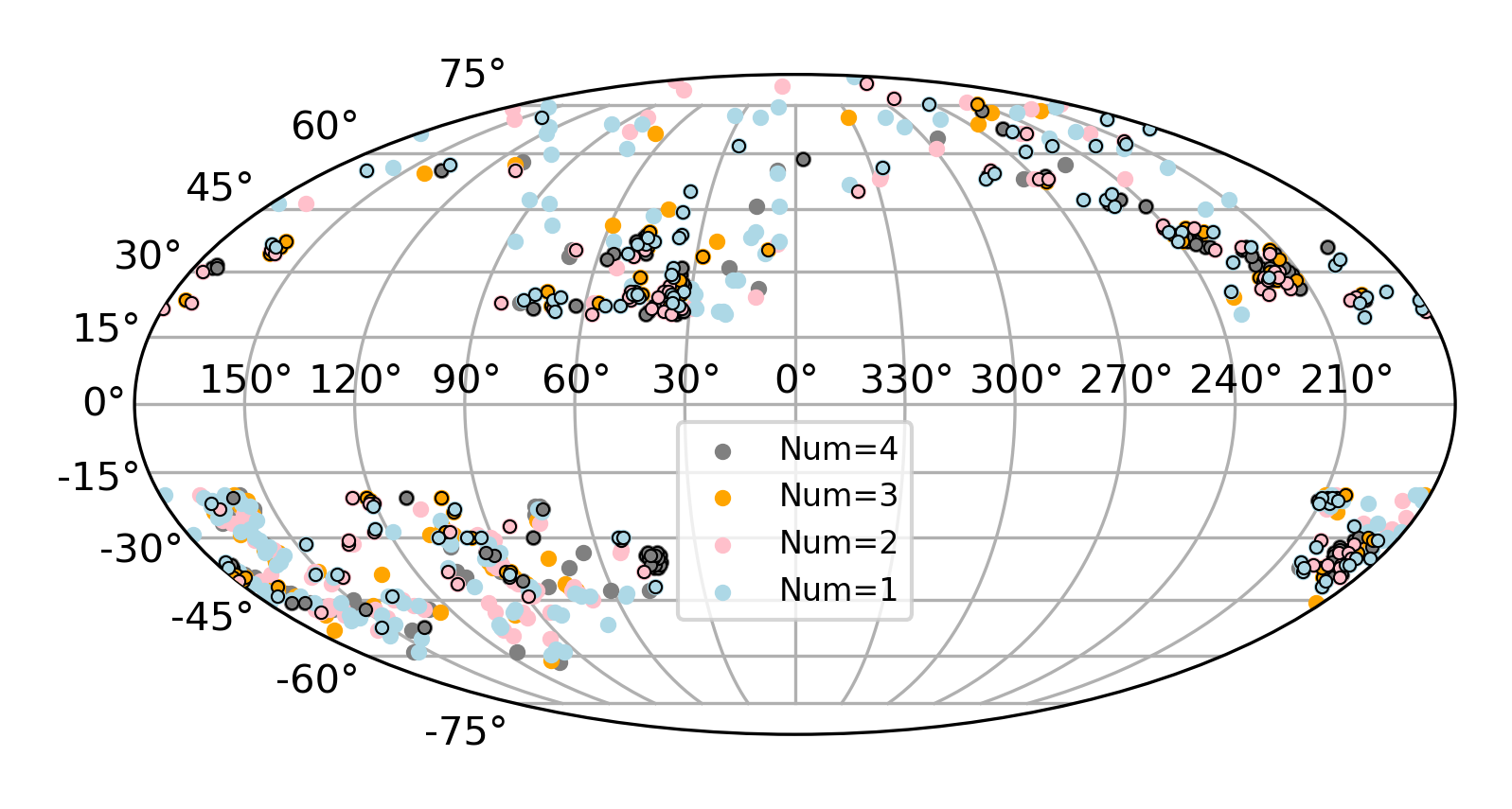}{0.7\textwidth}{}
            }
  \caption{Spatial distribution of the pixels eliminated in the four reddening maps in Galactic coordinates. 
  Gray, orange, pink, and light-blue dots represent pixels removed in four, three, two, and one maps, respectively. 
  The black circles represent pixels where dust clouds similar to those in Figure\,\ref{Fig7:samples_fgdust} are found.}
  \label{Fig8:cutdots}
\end{figure*}

We have checked the reddening-distance profiles of these pixels. About half (374) of them have dust clouds at large distances to the Galactic plane $Z$ (Figure \ref{Fig7:samples_fgdust}). For these pixels, there are lots of stars locating in front of the dust cloud. Thus the reddening values of these stars $E(B-V)_{\rm LAMOST}$ are much smaller than those from the reddening maps $E(B-V)_{\rm map}$, causing large values of Std$_{\rm fit}$. 
Figure \ref{Fig8:cutdots} shows the spatial distribution of all the discarded pixels for the four dust maps.
The number of rejected times are also marked. It must be admitted that not all pixels with associated dust cloud
are well eliminated in the four extinction maps. However, due to the small number of such pixels,  they will not affect the statistical results obtained in the following analyses.

The final spatial distributions of $E(B-V)_\mathrm{LAMOST}$ and $E(B-V)_\mathrm{map}$
for the SFD and \emph{Planck} maps are shown in Figure \ref{Fig9:LAMOSTandfourmaps_ebv}.
The differences ($E(B-V)_{\rm LAMOST} - E(B-V)_{\rm map}$  and $k$) are shown in Figure \ref{Fig10:deltaandk_fourmaps}
and Figure \ref{Fig11:hist_deltaandk_fourmaps}.
On one hand, the overall agreements between the $E(B-V)_\mathrm{LAMOST}$ and $E(B-V)_\mathrm{map}$ maps are 
excellent. On the other hand, the small yet significant discrepancies show clear spatially-dependent patterns. 
The patterns for the SFD and Planck2014-R maps are similar, while those for the Planck2014-Tau and Planck2019-Tau maps are similar.

\subsection{Precisions of the SFD and \emph{Planck} maps}\label{subsec:kfactors}

Before we discuss the possible factors that affect the spatial variations of $k$, 
we first check and validate reddening correction precisions of the SFD and \emph{Planck} maps.
As there are systematics of $E(B-V)$ values between the four reddening maps, we compare $E(G_{\rm BP}-G_{\rm RP})$ values here. 
For each pixel of each map, we estimate the dispersion of the differences between $E(G_{\rm BP}-G_{\rm RP})_{\rm LAMOST}$ and $E(G_{\rm BP}-G_{\rm RP})_{\rm map}$, where $E(G_{\rm BP}-G_{\rm RP})_{\rm map}$ is calculated using the corresponding reddening coefficient.
For each map, the median dispersion values are plotted against $E(G_{\rm BP}-G_{\rm RP})_{\rm LAMOST}$ 
in Figure \ref{Fig12:EBPRP_compare}. Dispersion values increase with the reddening values of the pixels. 
The curves of the four maps are very close, suggesting that 
the four reddening maps can achieve similar precisions in reddening correction. 
In the low-extinction region ($E(G_{\rm BP}-G_{\rm RP})$ $<$ 0.1\,mag), the Planck2014-Tau map is slightly worse. 
In the relatively high-extinction region, the Planck2019-Tau map is slightly better. 
The typical dispersion is 0.016, 0.022, and 0.045\,mag at $E(G_{\rm BP}-G_{\rm RP})_{\rm LAMOST} =$ 0.01, 0.1, and 0.7\,mag,
respectively. Note the dispersion values include measurement uncertainties of $E(G_{\rm BP}-G_{\rm RP})_{\rm LAMOST}$, 
therefore are upper limits of the reddening correction precisions.

\subsection{Properties correlated with the correction factor}

In Figure \ref{Fig10:deltaandk_fourmaps}, the spatial distributions of the correction factor \emph{k} show clear patterns. We explore here the possible origins of the patterns. 

\subsubsection{Reddening values}
\label{subsubsec:factorEBV}

The panels in the first row of Figure \ref{Fig13:k-EBV-area} plot the correction factor $k$ against reddening values $E(B-V)$ from the SFD and \emph{Planck} maps. All the pixels are divided into different bins 
according to their $E(B-V)$ values. The bin width is 0.01\,mag when $E(B-V) <$ 0.1\,mag, and 0.1\,mag otherwise.
The median values of $E(B-V)_{\rm map}$ and $k$ are over-plotted in red pluses and listed in Table \ref{Tab2:kandEBV}.

The overall relations between $k$ and $E(B-V)_{\rm map}$ are similar for the four reddening maps (particularly 
between the SFD and Planck2014-R maps, and between the Planck2014-Tau and Planck2019-Tau maps).
All show dips at $E(B-V)_{\rm map} \sim 0.03$ mag  when $E(B-V)_{\rm map}$ $<$ 0.1\,mag.
The trends become flat around 1 when $E(B-V)_{\rm map}$ $>$ 0.1\,mag, 
as expected considering that reddening-dependent coefficients are used (see Section\,\ref{subsec:calEBV}) to compute
the $E(B-V)_{\rm LAMOST}$ values and then the $k$ values.
The dips in low reddening regions are likely caused by the stronger over-estimation of reddening compared to 
that in the high reddening regions.

We have explored the relations between $k$ and $E(B-V)_{\rm map}$ for five different sky areas in the remaining rows of Figure \ref{Fig13:k-EBV-area}. For a given sky area, the relations are similar between the SFD and Planck2014-R maps, and between the Planck2014-Tau and Planck2019-Tau maps. However, for a given map, the relations are different between different sky areas, suggesting that reddening is not the main factor for the spatial variations of $k$.    

\subsubsection{Dust Temperature and spectral index}
\label{subsubsec:factor_dustTandbeta}

Since the SFD and \emph{Planck} maps are all based on the MBB model of dust thermal emissions, the correction factor $k$ could be correlated to the model parameters, i.e. dust temperature $T_{\rm d}$ and spectral index $\beta$. 
Note that  $T_{\rm d}$ and  $\beta$ parameters are usually fitted simultaneously and suffer strong 
anti-correlation in the MBB model (e.g., see Figure 7 of Planck2014). 
The SFD map adopted a constant value of $\beta$ = 2. While the Planck2014-R map was based on dust 
radiance ($\mathcal{R}$),  which is the integrated intensity  (See Equation\,9 in Planck2014)
and thus does not suffer from any degeneracy in the fit parameters. 
The possible effect of spectral index $\beta$ on the Planck2014-R map should be very weak.
Therefore, we first investigate the effect of  $T_{\rm d}$ on $k$ for the SFD and Planck2014-R maps.

The upper-left and upper-right panels of Figure \ref{Fig14:dustT_EBV_SFDandPLanck2014-R} plot $k$ as functions of $E(B-V)$ and $T_{\rm d}$ for the SFD and Planck2014-R maps, respectively. A strong dependence on $T_{\rm d}$
is found for the SFD map, particularly in the low extinction regions ($E(B-V) < 0.3$\,mag), where $k$ decreases when $T_{\rm d}$ increases. It suggests that the SFD map underestimates/overestimates reddening in low/high dust temperature regions. One possible explanation is that the SFD $T_{\rm d}$ map "underestimates/overestimates" $T_{\rm d}$ in high/low dust temperature regions. The result is consistent with \citet{peek_correction_2010} and \citet{lenz_new_2017}.

For the Planck2014-R map, the dependence on $E(B-V)$ and $T_{\rm d}$ is more clearly visible, in both the low and high extinction regions. It suggests that the Planck2014-R map underestimates/overestimates reddening in low/high dust temperature regions, and indicates that the Planck dust radiance map underestimates/overestimates the integrated intensity in low/high dust temperature regions.

We also investigate the dependence of $k$ on $T_{\rm d}$ for the Planck2014-Tau and Planck2019-Tau maps.
Very weak relations are found. However, a strong dependence on $\beta$ is found for the above two maps. 
To show the dependence more clearly, we first perform a normalization of the $k$ values with respect to 
the red lines in the top panels of Figure \ref{Fig13:k-EBV-area} to get rid of the effect of reddening. 
The normalized $k$ values are named $k^{'}$ hereafter. The upper-left panel of Figure \ref{Fig15:kfit_planck2014tau} shows $k^{'}$ as functions of $T_{\rm d}$ and  $\beta$ for the 
Planck2014-Tau map. We also divide the pixels into four different reddening ranges, and their results are 
plotted in other upper panels. It can be seen that for a given  $T_{\rm d}$, $k^{'}$ increases as  $\beta$
increases. Similar results are found for the Planck2019-Tau map, as shown in Figure \ref{Fig16:kfit_planck2019tau}.

\subsection{Fitting of the correction factor}
\label{subsec:kfit}

To provide corrections of the SFD and \emph{Planck} maps for regions outside the footprints in the current work, 
we fit the correction factor $k$ as functions of $E(B-V)$ and $T_{\rm d}$ for the SFD and Planck2014-R maps, 
and the normalized correction factor $k^{'}$ as functions of $T_{\rm d}$ and  $\beta$ for 
the  Planck2014-Tau and Planck2019-Tau maps. 

Two-dimensional fourth-order polynomials of 15 free parameters are adopted in the fitting with
the {\sc{Python}} procedure {SciPy}. The resultant coefficients are listed in Table \ref{Tab3:kfit}.
The fitting residuals are shown in Figure \ref{Fig14:dustT_EBV_SFDandPLanck2014-R} for the SFD and Planck2014-R maps, Figure \ref{Fig15:kfit_planck2014tau} for the Planck2014-Tau map, 
and Figure \ref{Fig16:kfit_planck2019tau} for the Planck2019-Tau map, respectively. 
One can see that global trends with $E(B-V)$, $T_{\rm d}$ or  $\beta$ are successfully removed. 

Figure \ref{Fig17:resk-healpix} shows spatial distributions of $k$/$k^{'}$,  $k_{\rm fit}$/$k^{'}_{\rm fit}$,
and $\Delta k$/$\Delta k^{'}$ for the SFD and Planck maps. 
The fluctuations are smaller in the residual maps.
However, the fitting residuals still show similar spatial patterns, despite the fact that they 
show no dependence on $E(B-V)$, $T_d$ or $\beta$. Figure \ref{Fig18:deltak_correlation} compares 
fitting residuals between different maps. The correlation coefficients are between 0.63 -- 0.90. 
The result indicates that there are likely other factors (e.g., dust sizes/reddening laws) 
that contribute the spatial variations of the correction factors. Such possibilities will be 
explored in the future work.

\section{Summary}\label{sec:Sum}

\begin{figure*}
  \gridline{\fig{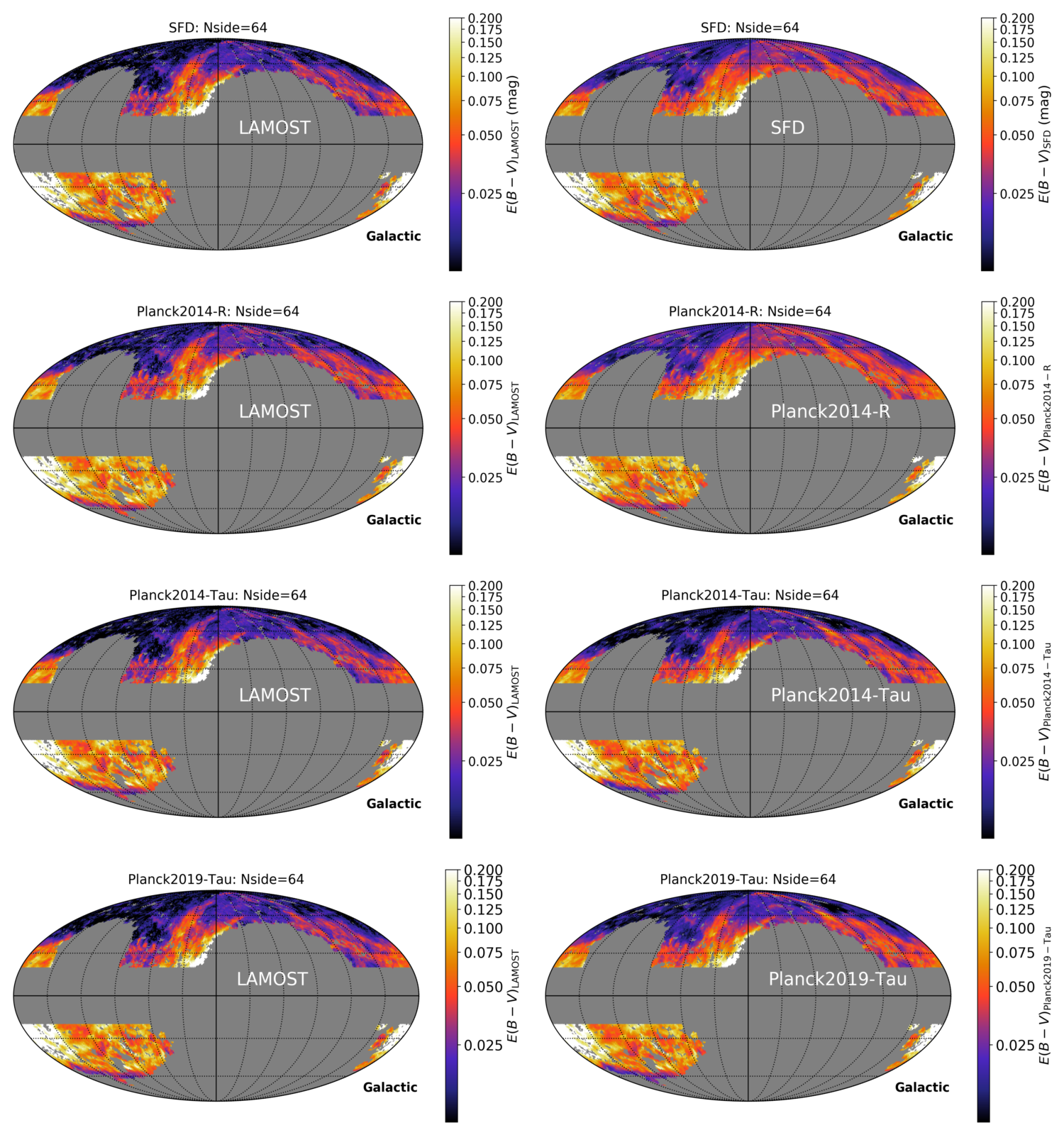}{0.9\textwidth}{}
            }
  \caption{All-sky distributions of $E(B-V)_\mathrm{LAMOST}$ (left) and $E(B-V)_\mathrm{map}$ (right). 
  From top to bottom are for the SFD, Planck2014-R, Planck2014-Tau and Planck2019-Tau maps, respectively. 
  Note the four maps in the left are different due to different reddening coefficients used.}
  \label{Fig9:LAMOSTandfourmaps_ebv}
\end{figure*}

\begin{figure*}
  \gridline{\fig{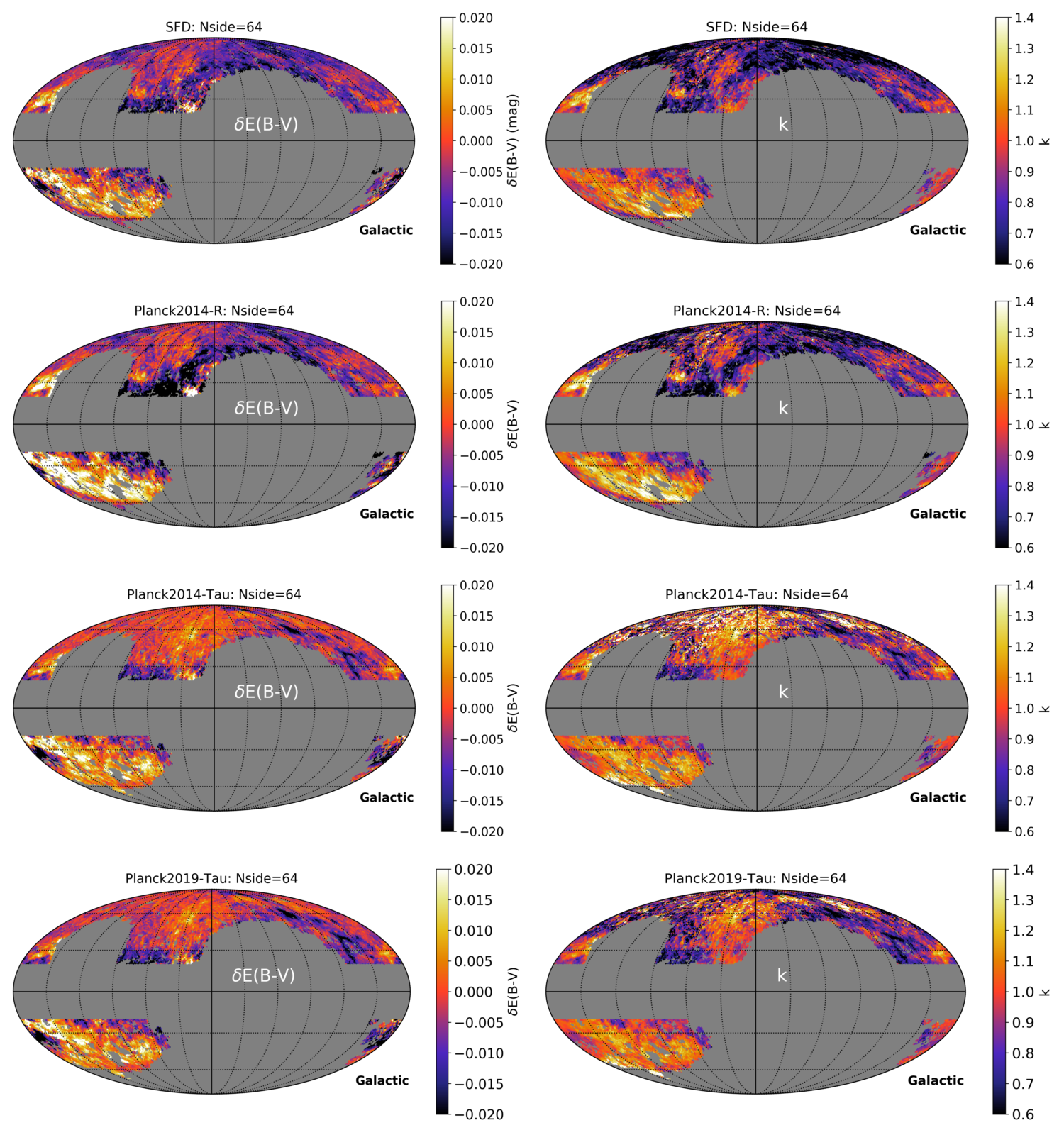}{0.9\textwidth}{}
            }
  \caption{All-sky distributions of$E(B-V)_{\rm LAMOST} - E(B-V)_{\rm map}$ (left) and $k$ (right).
  From top to bottom are for the SFD, Planck2014-R, Planck2014-Tau and Planck2019-Tau maps, respectively.}
  \label{Fig10:deltaandk_fourmaps}
\end{figure*}

\begin{figure*}
  \gridline{\fig{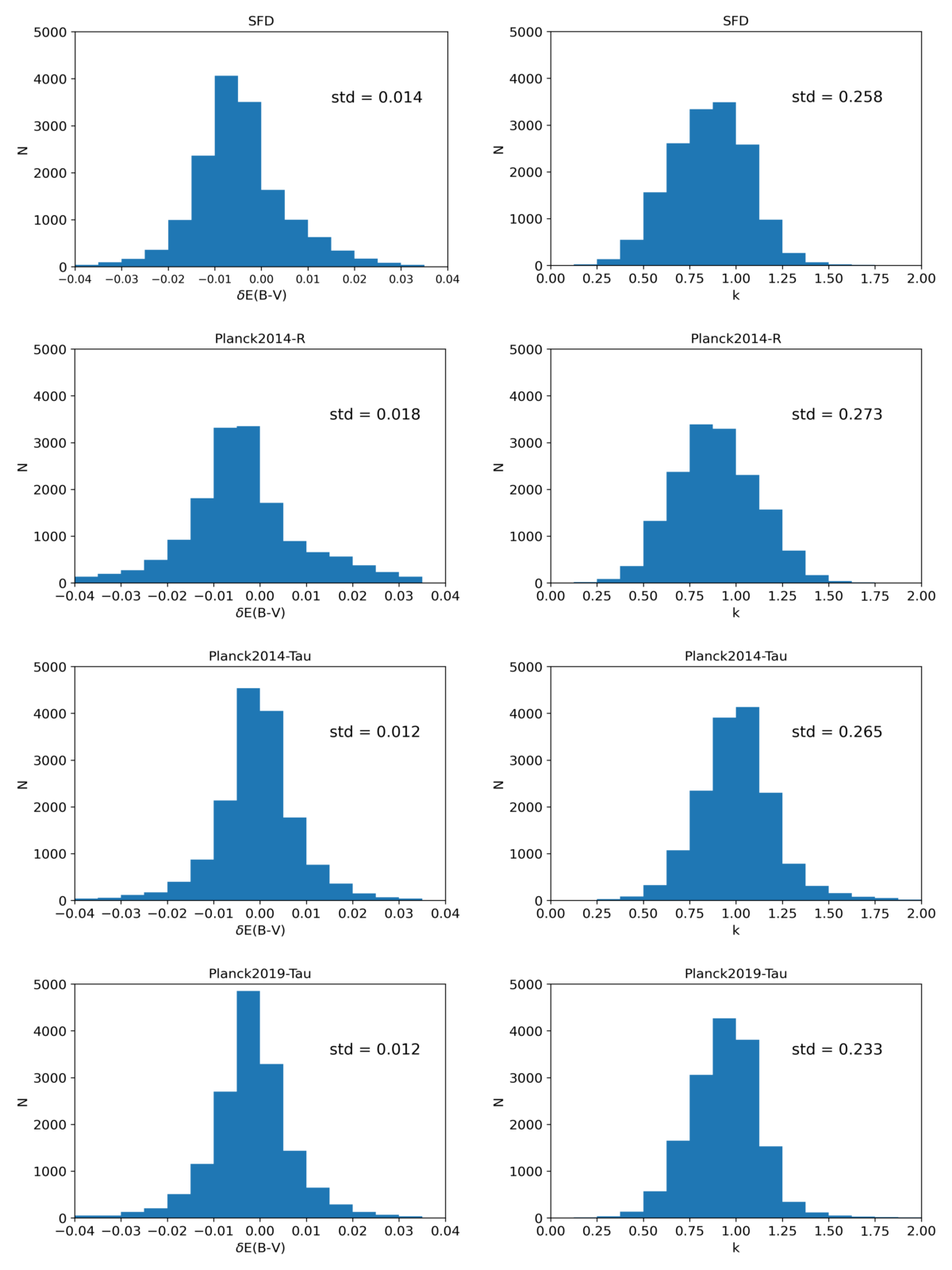}{0.9\textwidth}{}
            }
  \caption{Histogram distributions of $E(B-V)_{\rm LAMOST} - E(B-V)_{\rm map}$ (left) and $k$ (right).
  From top to bottom are for the SFD, Planck2014-R, Planck2014-Tau and Planck2019-Tau maps, respectively.}
  \label{Fig11:hist_deltaandk_fourmaps}
\end{figure*}

\begin{figure*}
  \gridline{\fig{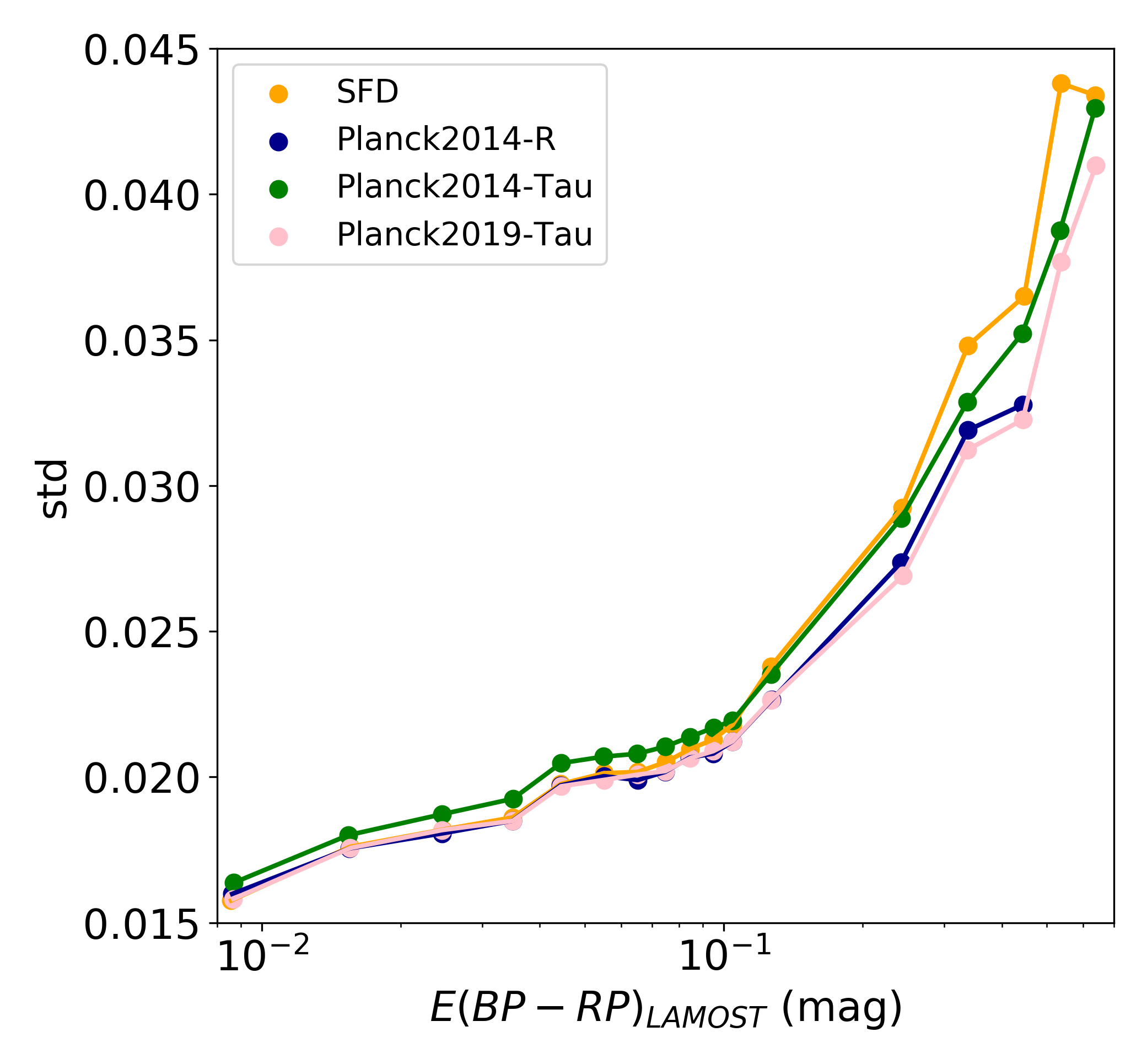}{0.7\textwidth}{}
            }
  \caption{Typical dispersion values of the differences between $E(G_{\rm BP}-G_{\rm RP})_{\rm LAMOST}$ and $E(G_{\rm BP}-G_{\rm RP})_{\rm map}$
  plotted against $E(G_{\rm BP}-G_{\rm RP})_{\rm LAMOST}$. The orange, blue, green and pink dots represent results for the SFD, Planck2014-R, Planck2014-Tau and Planck2019-Tau dust maps, respectively.}
  \label{Fig12:EBPRP_compare}
\end{figure*}

\begin{figure*}
  \gridline{\fig{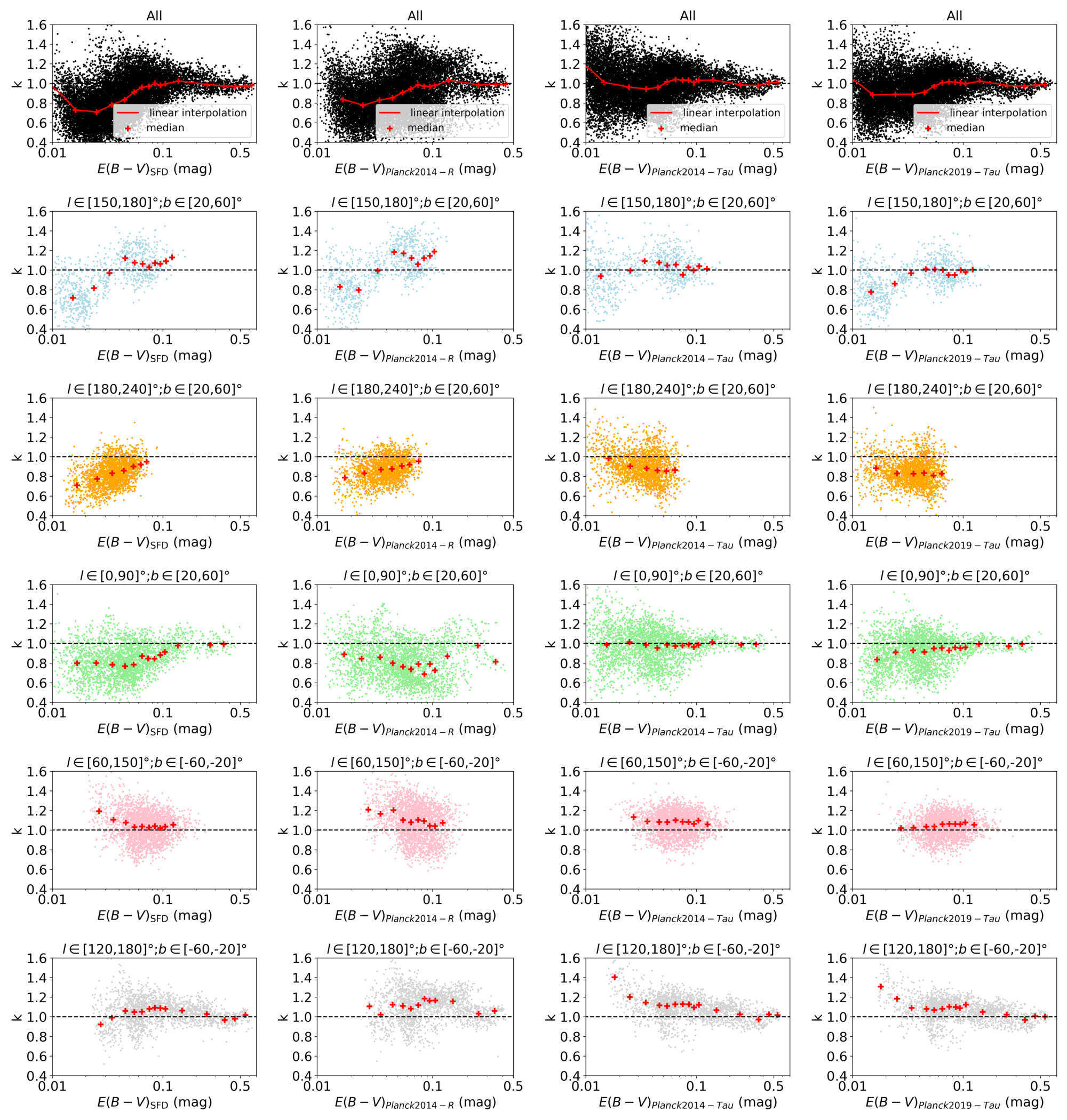}{1.0\textwidth}{}
            }
  \caption{Correction factor k as a function of $E(B-V)_\mathrm{map}$ in different sky areas. From left to right are for the SFD, Planck2014-R, Planck2014-Tau, and Planck2019-Tau maps, respectively.
  For each panel, the sky area is labelled on the top, and the red crosses mark the median values. The trends with 
  $E(B-V)_\mathrm{map}$ are different between different areas for all the maps.
  }
  \label{Fig13:k-EBV-area}
\end{figure*}

\begin{figure*}
  \gridline{\fig{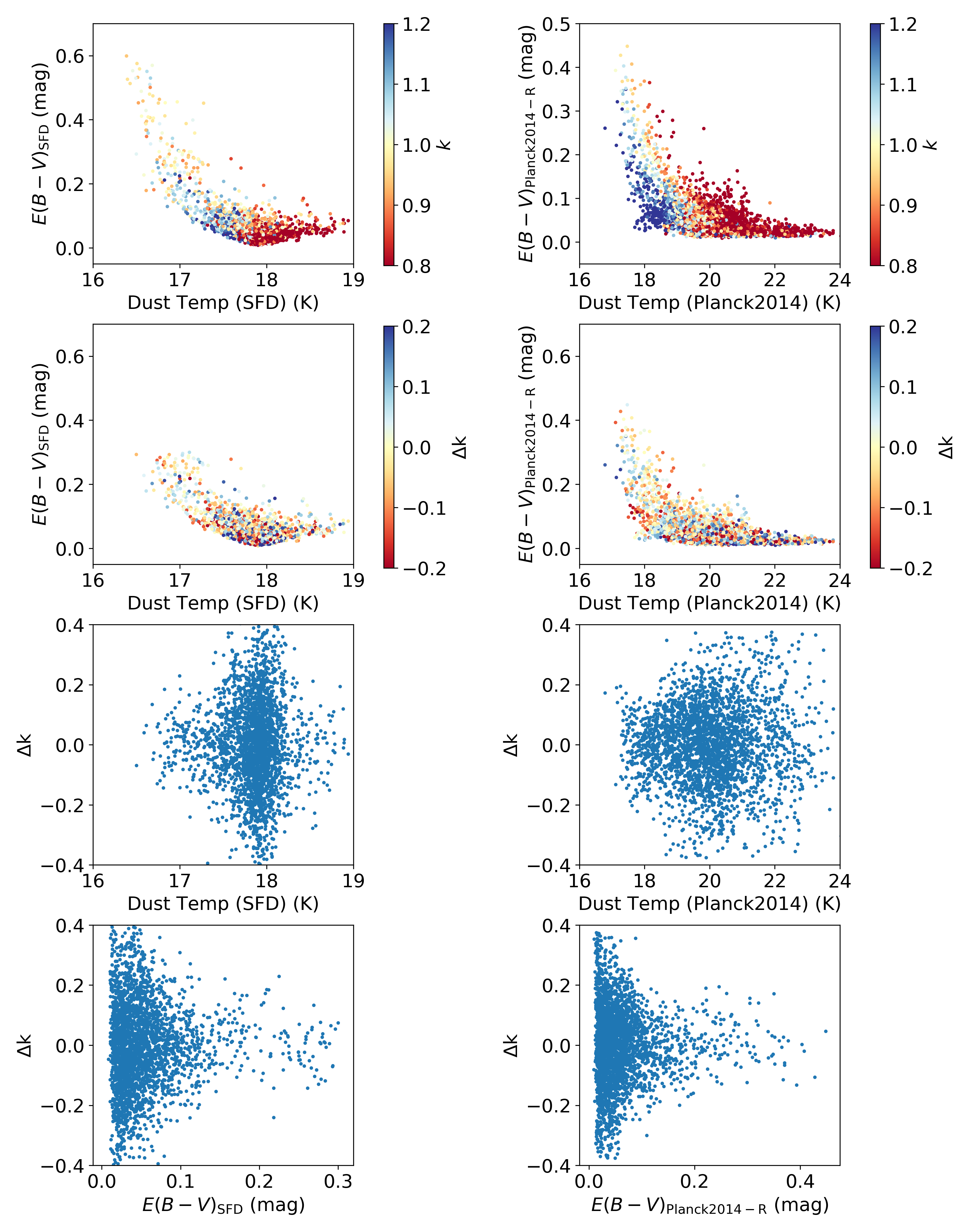}{0.7\textwidth}{}
            }
  \caption{Top row: dependence of $k$ on dust temperature and reddening for the SFD and Planck2014-R maps. 
  Second row: fitting residuals. 
  Third row: fitting residuals as a function of dust temperature. Bottom row: fitting residuals as a function of reddening. One-fifth of all pixels are randomly selected to plot the results.}
  \label{Fig14:dustT_EBV_SFDandPLanck2014-R}
\end{figure*}

\begin{figure*}
  \gridline{\fig{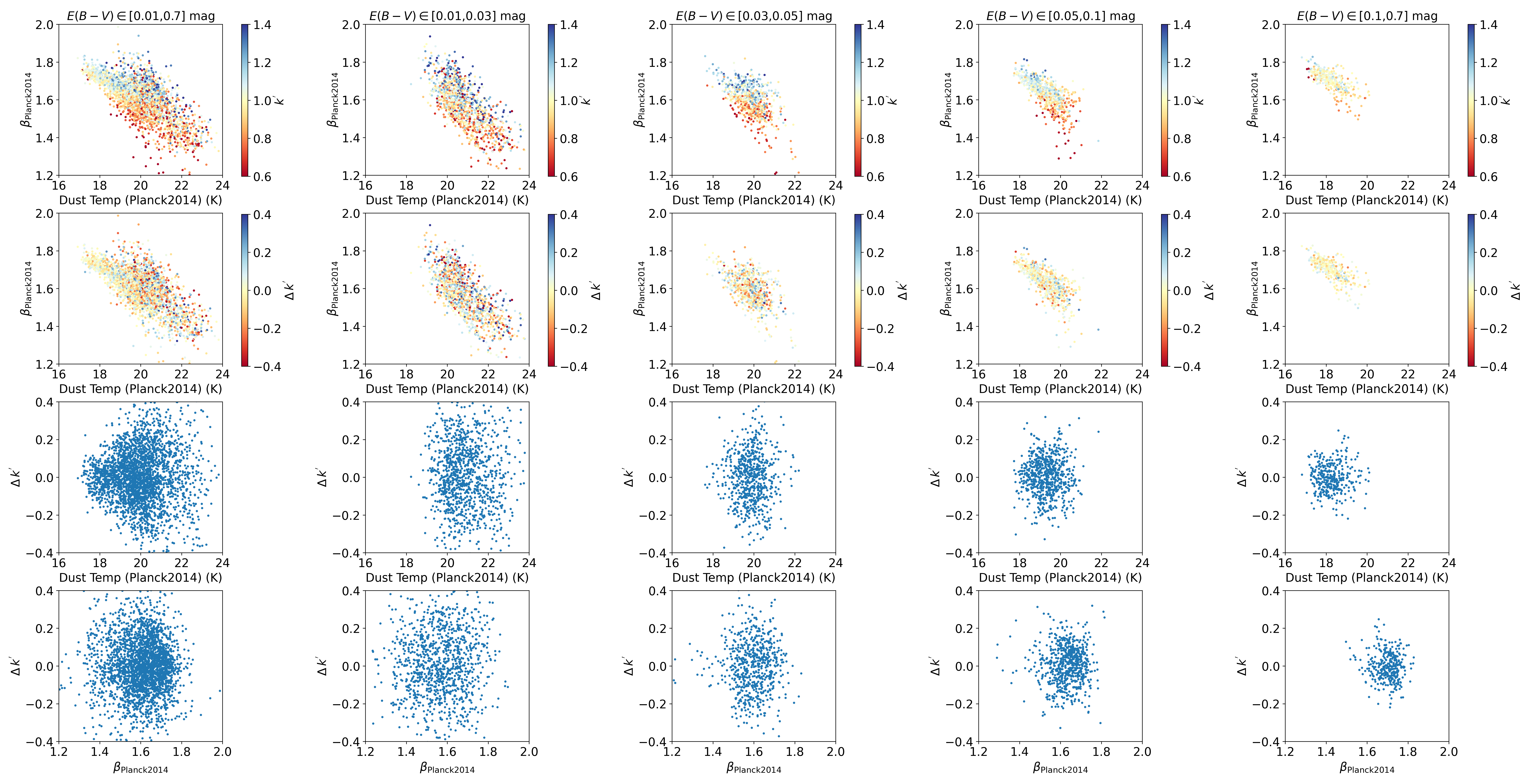}{1.0\textwidth}{}
            }
  \caption{Top row: dependence of $k^{'}$ on dust temperature and spectral index for the Planck2014-Tau map. 
  Second row: Fitting residuals. Third row: fitting residuals as a function of dust temperature. Bottom row: fitting residuals as a function of spectral index. The first column shows the results of all pixels. The other four columns show the results of pixels in four different reddening ranges. One-fifth of all pixels are randomly selected to plot the results.}
  \label{Fig15:kfit_planck2014tau}
\end{figure*}

\begin{figure*}
  \gridline{\fig{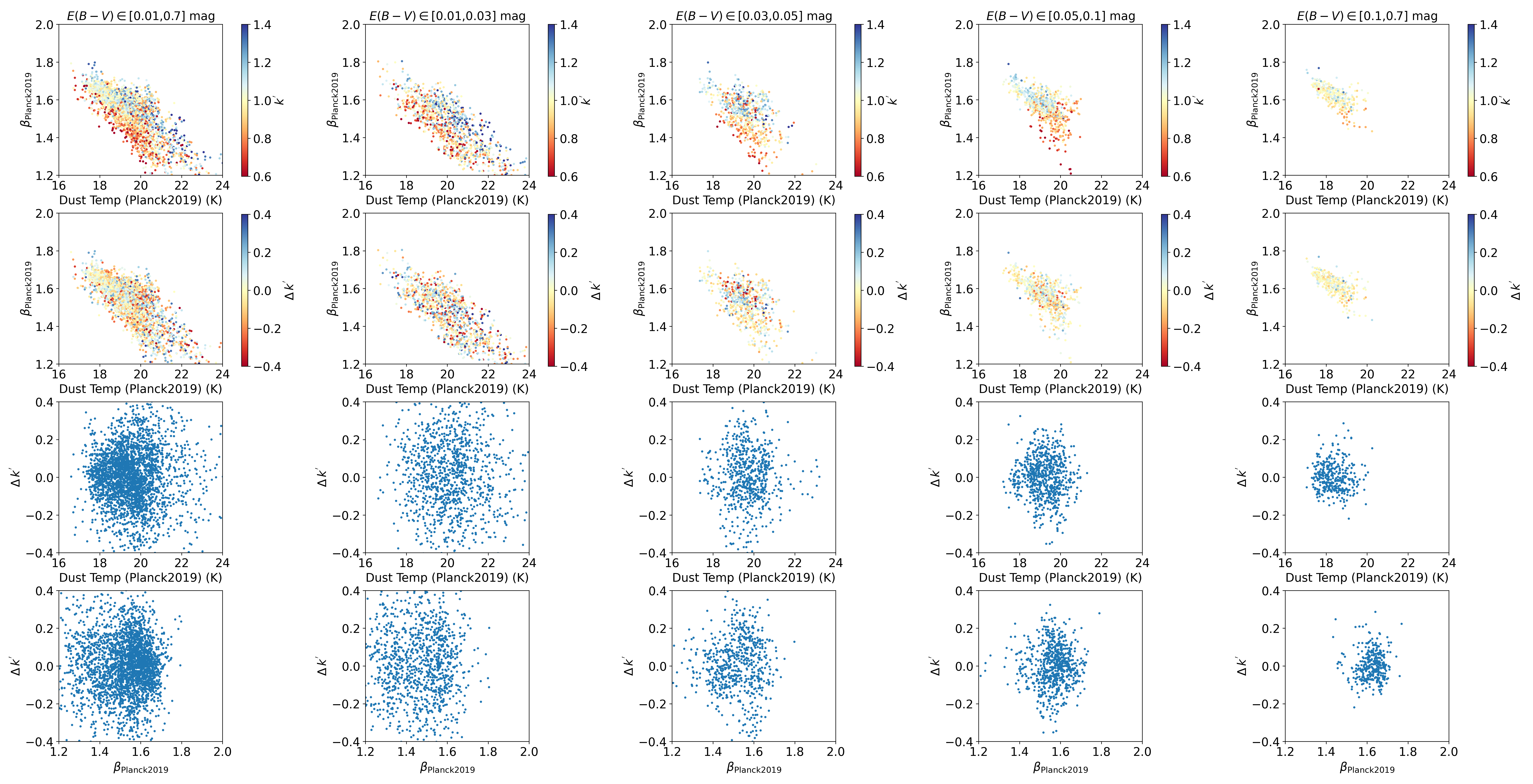}{1.0\textwidth}{}
            }
  \caption{Same as Figure\,\ref{Fig15:kfit_planck2014tau} but for the Planck2019-Tau map.}
  \label{Fig16:kfit_planck2019tau}
\end{figure*}

Combining high-precision LAMOST DR5 spectroscopic data and  \emph{Gaia} DR2  photometric data, 
we have calculated the color excess $E(G_{\rm BP}-G_{\rm RP})$ for about two million well selected 
middle and high Galactic latitude stars by using the star-pair technique. 
With empirical temperature- and reddening-dependent coefficients, values of $E(B-V)_\mathrm{LAMOST}$ are further obtained to check and correct the SFD, Planck2014-R, Planck2014-Tau, and Planck2019-Tau reddening maps. By comparing $E(B-V)_\mathrm{LAMOST}$ with $E(B-V)_\mathrm{map}$,  the following results are found:

\begin{enumerate}

\item On one hand, the overall agreements between the $E(B-V)_\mathrm{LAMOST}$ and $E(B-V)_\mathrm{map}$ maps 
are excellent (Figure \ref{Fig9:LAMOSTandfourmaps_ebv}). On the other hand, the small yet significant discrepancies show clear spatially-dependent patterns (Figure \ref{Fig10:deltaandk_fourmaps}). 
The patterns for the SFD and Planck2014-R maps are similar, while those for the Planck2014-Tau and Planck2019-Tau maps are similar.

\item The four reddening maps can achieve similar precisions in reddening correction (Figure \ref{Fig12:EBPRP_compare}). 
In the low-extinction region ($E(G_{\rm BP}-G_{\rm RP})$ $<$ 0.1\,mag), the Planck2014-Tau map is slightly worse. 
In the relatively high-extinction region, the Planck2019-Tau map is slightly better. 

\item For a given sky area, the  $k$ and $E(B-V)_{\rm map}$ relations are similar between the SFD and Planck2014-R maps, and between the Planck2014-Tau and Planck2019-Tau maps. However, for a given map, the relations are different among different sky areas (Figure \ref{Fig13:k-EBV-area}), suggesting that reddening is not the main factor for the spatial variations of $k$.   

\item For the SFD and Planck2014-R maps, the dependence of $k$ on $E(B-V)$ and $T_{\rm d}$ is clearly visible (Figure \ref{Fig14:dustT_EBV_SFDandPLanck2014-R}). $k$ decreases when $T_{\rm d}$ increases. It suggests that the two maps underestimates/overestimates reddening in low/high dust temperature regions. One possible explanation is that the SFD $T_{\rm d}$ map "underestimates/overestimates" $T_{\rm d}$ in high/low dust temperature regions, consistent with \citet{peek_correction_2010} and \citet{lenz_new_2017}. While the Planck dust radiance map possibly underestimates/overestimates the integrated intensity in low/high dust temperature regions.

\item For the Planck2014-Tau and Planck2019-Tau maps, the dependence of $k$ on $T_{\rm d}$ is very weak, but very strong on $\beta$. For a given  $T_{\rm d}$, the normalized correction factor $k^{'}$ increases as $\beta$ increases (Figure \ref{Fig15:kfit_planck2014tau} and Figure \ref{Fig16:kfit_planck2019tau}).

\end{enumerate}

\begin{figure*}
  \gridline{\fig{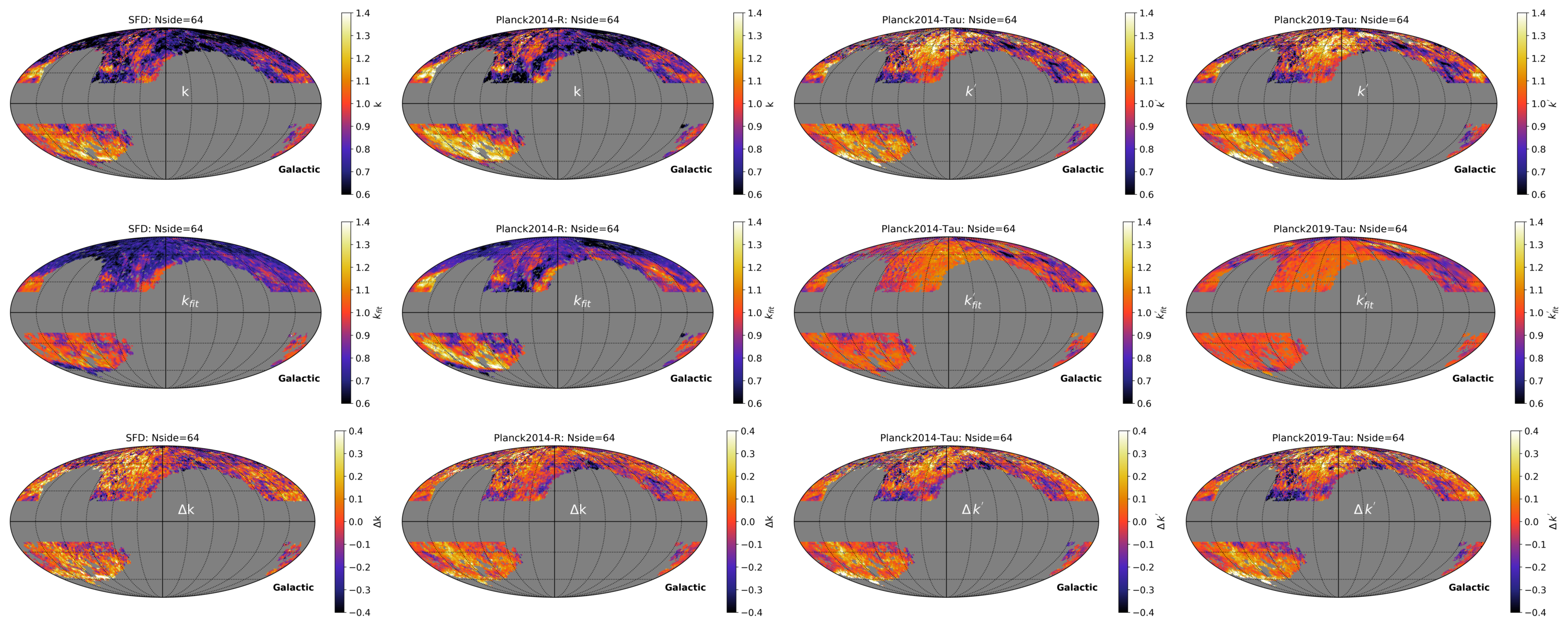}{1.0\textwidth}{}
            }
  \caption{All-sky maps of k (or $k^{'}$) (top), $k_{fit}$ (or $k_{fit}^{'}$) (middle) and fitting residual $\Delta\,k$ (or $\Delta\,k^{'}$) (bottom) for different reddening maps. From left to right are for the SFD, Planck2014-R, Planck2014-Tau, and Planck2019-Tau maps, respectively.}
  \label{Fig17:resk-healpix}
\end{figure*}

\begin{figure*}
  \gridline{\fig{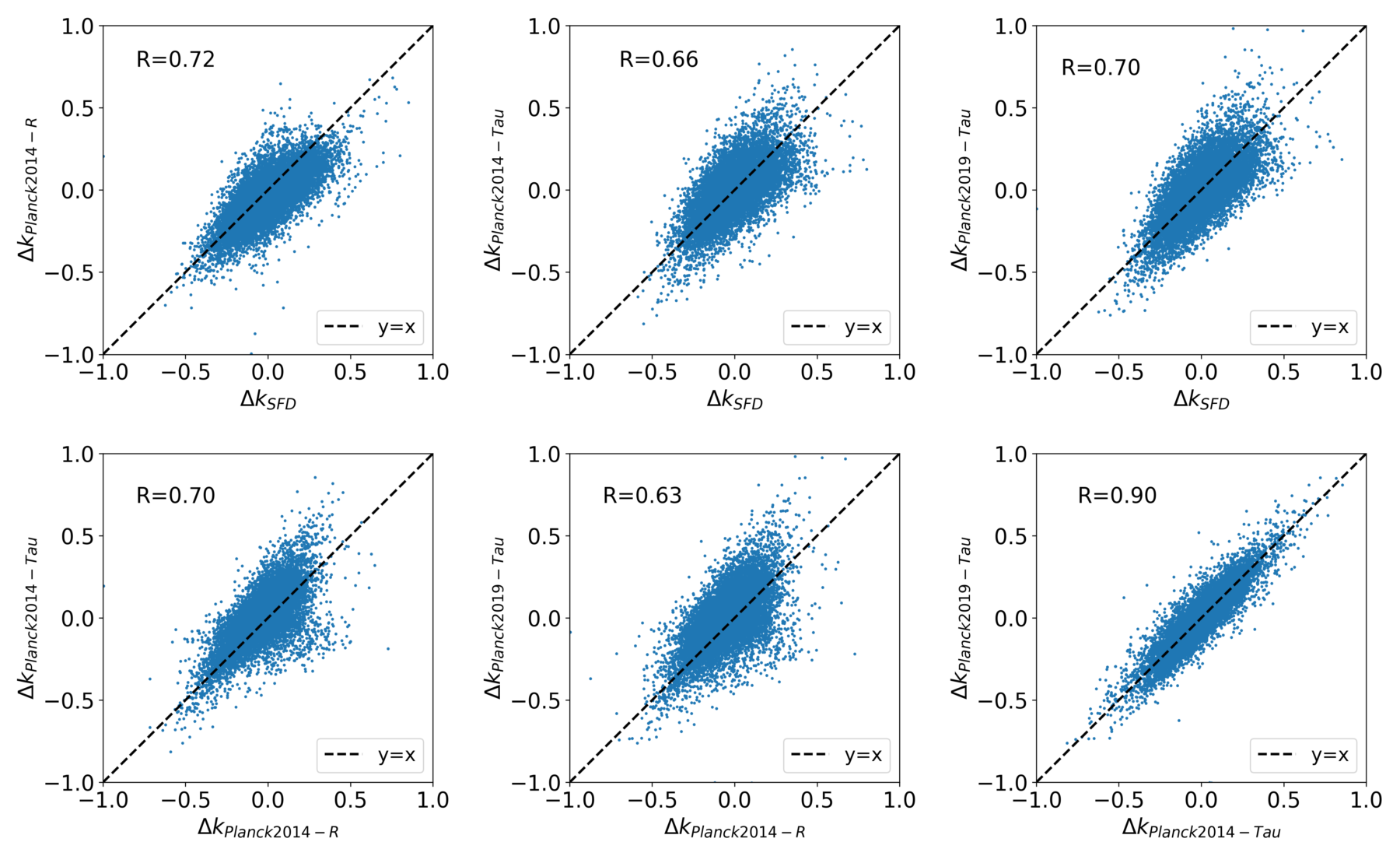}{1.0\textwidth}{}
            }
  \caption{Comparisons between the fitting residuals $\Delta\,k$ (or $\Delta\,k^{'}$) of the SFD, Planck2014-R, Planck2014-Tau, and Planck2019-Tau maps. The correlation coefficients $R$ are labeled.}
  \label{Fig18:deltak_correlation}
\end{figure*}

The $k$ maps and their errors are publicly available\footnote{http://paperdata.china-vo.org/Dustmaps-correction/extinction-maps-correction.zip} and can be used to perform corrections of the SFD and \emph{Planck} maps.
For regions outside the footprints in the current work, relations of $k$ as functions of $E(B-V)$ and $T_{\rm d}$ for the SFD and Planck2014-R maps, and  $k^{'}$ as functions of $T_{\rm d}$ and  $\beta$ for 
the Planck2014-Tau and Planck2019-Tau maps, can be used to correct global trends with $E(B-V)$, $T_{\rm d}$ or  $\beta$ 
to some extent (Table \ref{Tab3:kfit} and Figure \ref{Fig17:resk-healpix}). 
It should be noticed that, applications of the empirical correction relations are limited by the range of $E(B-V)$ for fitting.
For the convenience of use, a python routine is provided\footnote{https://github.com/qy-sunyang/Extinction-Maps-Correction} for such purposes.
However, the fitting residuals between different maps still show similar spatial patterns and 
good correlations (Figure \ref{Fig18:deltak_correlation}), 
indicating that there are likely other factors (e.g., dust sizes/reddening laws) 
that contribute the spatial variations of the correction factors. 
Such possibilities will be explored in the future work.

Our results provide important clues for the further improvement of the Galactic all-sky extinction maps and lay an important foundation for the accurate extinction correction in the era of precision astronomy.

\acknowledgments
We acknowledge the referee for his/her valuable comments to improve the clarity and quality 
of the manuscript.
This work is supported by the National Key Basic R\&D Program of China via 2019YFA0405503, the National Natural Science Foundation of China through the projects NSFC 12173007, 12173034 and 11603002, and Beijing Normal University grant No. 310232102. 
We acknowledge the science research grants from the China Manned Space Project with NO. CMS-CSST-2021-A08 and CMS-CSST-2021-A09.
This work has made use of data from the European Space Agency (ESA) mission Gaia (\url{https://www.cosmos.esa.int/gaia}), processed by the Gaia Data Processing and Analysis Consortium (DPAC, \url{https:// www.cosmos.esa.int/web/gaia/dpac/consortium}). Funding for the DPAC has been provided by national institutions, in particular the institutions participating in the Gaia Multilateral Agreement. Guoshoujing Telescope (the Large Sky Area Multi-Object Fiber Spectroscopic Telescope LAMOST) is a National Major Scientific Project built by the Chinese Academy of Sciences. Funding for the project has been provided by the National Development and Reform Commission. LAMOST is operated and managed by the National Astronomical Observatories, Chinese Academy of Sciences. 

\bibliography{SFD}{}
\bibliographystyle{aasjournal}

\clearpage
\begin{deluxetable*}{cccccccc}
  \tablewidth{0pt}
  \tablenum{1}
  \tablecaption{Temperature- and reddening-dependent reddening coefficients of the Gaia DR2 passbands for different reddening maps. The function form is: $R = C_{0}+C_{1}y+C_{2}y^2+C_{3}x+C_{4}xy+C_{5}x^2$, where x is stellar temperature and y is $E(B-V)_{map}$.
  \label{Tab1:R_BPRPandBPG}}
  \tablehead{
  \colhead{Band} & \colhead{Extinction Map} & \colhead{$C_0$} & \colhead{$C_1$} & \colhead{$C_2$} & \colhead{$C_3$}   & \colhead{$C_4$}    & \colhead{$C_5$}} 
  \startdata
    {$R_{G_{\rm BP}-G_{\rm RP}}$} & SFD & 1.097E+00 & $-$2.415E$-$01 & 2.437E$-$01 & $-$7.219E$-$06 & $-$4.504E$-$05 & 7.716E$-$09 \\
    {} & Planck2014-R    & 5.559E$-$01 & 2.738E+00 & $-$3.006E+00 & 5.830E$-$05 & $-$3.468E$-$05 & 3.744E$-$09 \\
    {} &Planck2014-Tau  & 1.008E+00 & $-$1.445E$-$03 & $-$2.799E$-$01 & 3.136E$-$05 & $-$1.893E$-$05 & 4.083E$-$09 \\
    {} &Planck2019-Tau  & 1.044E+00 & $-$6.634E$-$02 & $-$2.316E$-$01 & 1.761E$-$05 & $-$2.670E$-$06 & 4.432E$-$09 \\
    \hline
     {$R_{G_{\rm BP}-G}$} & SFD & 9.040E$-$01 & $-$2.210E$-$01 & 1.750E$-$01 & $-$6.410E$-$05 & $-$1.150E-06 & 3.520E$-$09 \\
     {} & Planck2014-R    & 5.450E$-$01 & 1.260E+00 & $-$1.360E+00 & 7.520E$-$06 & 9.440E$-$06 & $-$2.620E$-$09 \\
     {} & Planck2014-Tau  & 8.560E$-$01 & $-$7.380E$-$02 & $-$1.360E$-$01 & $-$4.310E$-$05 & 1.570E$-$05 & 1.280E$-$09 \\
     {} & Planck2019-Tau  & 8.510E$-$01 & $-$1.260E$-$01 & $-$1.080E$-$01 & $-$4.390E$-$05 & 2.720E$-$05 & 1.030E$-$09 \\
  \enddata
\end{deluxetable*}

\begin{deluxetable*}{ccccccccBcccccccc}
  \tablewidth{0pt}
  \tablenum{2}
  \tablecaption{
  Correction factor $k$ as a function of $E(B-V)_{\rm map}$ for the SFD and Planck maps.
    \label{Tab2:kandEBV}}
  \tablehead{
  \multicolumn{2}{c}{SFD}  & \multicolumn{2}{c}{Planck2014-R} & \multicolumn{2}{c}{Planck2014-Tau} & \multicolumn{2}{c}{Planck2019-Tau} &\\ 
  \colhead{$E(B-V)$}  & \colhead{$k_{median}$} & \colhead{$E(B-V)$} & \colhead{$k_{median}$} & \colhead{$E(B-V)$} & \colhead{$k_{median}$} & \colhead{$E(B-V)$} & \colhead{$k_{median}$} } 
  \startdata
    0.009  & 0.997 & 0.017        & 0.835           & 0.009          & 1.216 & 0.009          & 1.065 \\
    0.016  & 0.728 & 0.025        & 0.777           & 0.015          & 1.010 & 0.015          & 0.884 \\
    0.025  & 0.710 & 0.035        & 0.830           & 0.025          & 0.962 & 0.024          & 0.890 \\
    0.034  & 0.783 & 0.045        & 0.854           & 0.035          & 0.944 & 0.035          & 0.890 \\
    0.045  & 0.834 & 0.055        & 0.909           & 0.045          & 0.960 & 0.045          & 0.911 \\
    0.055  & 0.913 & 0.065        & 0.944           & 0.054          & 1.017 & 0.055          & 0.975 \\
    0.065  & 0.962 & 0.075        & 0.985           & 0.065          & 1.041 & 0.065          & 1.008 \\
    0.075  & 0.971 & 0.085        & 0.972           & 0.075          & 1.033 & 0.074          & 1.017 \\
    0.084  & 1.002 & 0.095        & 0.969           & 0.085          & 1.035 & 0.085          & 1.013 \\
    0.095  & 0.981 & 0.105        & 0.980           & 0.095          & 1.005 & 0.094          & 1.006 \\
    0.105  & 0.997 & 0.136        & 1.029           & 0.105          & 1.032 & 0.105          & 1.000 \\
    0.138  & 1.024 & 0.249        & 0.985           & 0.141          & 1.035 & 0.141          & 1.023 \\
    0.252  & 0.989 & 0.350        & 0.992           & 0.251          & 0.984 & 0.251          & 0.978 \\
    0.355  & 0.976 & 0.427        & 0.991           & 0.362          & 0.980 & 0.358          & 0.972 \\
    0.450  & 0.969 &              &                 & 0.452          & 1.007 & 0.449          & 0.984 \\
    0.558  & 0.975 &              &                 & 0.539          & 1.010 & 0.549          & 0.990 \\
    0.632  & 0.980 &              &                 &                &       &                &       \\
           &       &              &                 &                &       &                & \\     
  \enddata
\end{deluxetable*}

\begin{deluxetable*}{ccccc}
  \tablewidth{0pt}
  \tablenum{3}
  \tablecaption{\explain{Table 3 in the first-version paper is wrong, here we updated it to the correct version.}
    Fitting coefficients of $k$ and $k^{'}$. The function form is: $z= ax^4 + by^4 + cyx^3 + dxy^3 + ex^2y^2 + fx^3 + gy^3 + hyx^2 + ixy^2 + jx^2 + ky^2 + lxy + mx + ny + o$. For the SFD and Planck2014-R maps, $z$ is $k$, $x$ is $T_{d}$, and $y$ is $E(B-V)$. For the Planck2014-Tau and Planck2019-Tau maps, 
    $z$ is $k^{'}$, $x$ is  $T_{d}$, and $y$ is  $\beta$.\\
    \label{Tab3:kfit}}
  \tablehead{
  \colhead{}
    & \colhead{SFD}        & \colhead{Planck2014-R} & \colhead{Planck2014-Tau} & \colhead{Planck2019-Tau} }
  \startdata
a & -0.0096664     & 0.0007690    & 0.0003437      & -0.0011918     \\
b & 1103.8527321   & -195.2549470 & 25.0345231     & -29.6874082    \\
c & 3.6652110      & 0.0578553    & 0.0308794      & -0.0822918     \\
d & 623.1533468    & -15.6974711  & 7.5020031      & -11.6586509    \\
e & 87.1827827     & 1.7125698    & 0.7068647      & -1.7087918     \\
f & -11592.2395358 & 454.3115073  & -309.0274163   & 399.8471526    \\
g & 0.3264288      & -0.0712684   & -0.0647263     & 0.2236456      \\
h & -219.0927722   & -4.2123580   & -3.8361633     & 10.0293453     \\
i & -3365.6297663  & -57.9146527  & -61.9580752    & 118.9355843    \\
j & 2.4443323      & 2.4844060    & 4.3976829      & -14.3621849    \\
k & 32345.3778957  & 440.2730008  & 1331.7124722   & -2046.8604171  \\
l & 4350.8798470   & 96.5155018   & 165.9610933    & -375.9821150   \\
m & -180.9410685   & -38.5440982  & -133.0115411   & 380.1986985    \\
n & -28699.9073759 & -707.2919166 & -2420.6173146  & 4497.9132419   \\
o & 1576.5430626   & 225.1130841  & 1513.4808280   & -3575.0589374  \\
  \enddata
  \end{deluxetable*}


\end{document}